\documentclass[pra,onecolumn,showpacs,superscriptaddress,amsmath,amssymb]{revtex4} 
\usepackage{graphicx,bm} 

\usepackage{amsmath}	
\usepackage{mathrsfs}%
\begin{document} 

\title{Dissipative hydrodynamic equation of a ferromagnetic Bose-Einstein 
condensate: Analogy to magnetization dynamics in conducting ferromagnets} 
\author{Kazue Kudo} 
\affiliation{Division of Advanced Sciences, Ochadai Academic Production, 
Ochanomizu University, 2-1-1 Ohtsuka, Bunkyo-ku, Tokyo 112-8610, Japan} 
\author{Yuki Kawaguchi} 
\affiliation{Department of Physics, University of Tokyo, 7-3-1 Hongo,
Bunkyo-ku, Tokyo 113-0033, Japan} 
 
\date{\today} 
\begin{abstract} 
 The hydrodynamic equation of a spinor Bose-Einstein condensate (BEC) gives a
 simple description of spin dynamics in the condensate. We introduce the
 hydrodynamic equation of a ferromagnetic BEC with dissipation 
 originating from the energy dissipation of the condensate. The dissipative
 hydrodynamic equation has the same form as an extended
 Landau-Lifshitz-Gilbert (LLG) equation, which describes the
 magnetization dynamics of conducting ferromagnets in which localized 
 magnetization interacts with spin-polarized
 currents.  Employing the dissipative
 hydrodynamic equation, we demonstrate the
 magnetic domain pattern dynamics of a ferromagnetic BEC in 
 the presence and absence of a current of particles, and discuss the
 effects of the current on domain pattern formation.
 We also discuss the characteristic lengths of domain patterns that have
 domain walls with and without finite magnetization.
\end{abstract} 
\maketitle 
 
\section{Introduction}

A particular feature of superfluids 
and superconductors with spin degrees of freedom,
such as superfluid Helium three, {\it p}-wave superconductors, and
spinor Bose-Einstein condensates (BECs) of ultra-cold atoms,  
is that they support the non-dissipative flow of spins, or spin
supercurrent~\cite{Leggett, Borovik}. 
In such systems, the condensed state is described with a multi-component
order parameter;  
the supercurrent of the particles in each spin state is proportional to
the phase gradient of the corresponding component of the order parameter; 
hence, the gradient of the relative phase of the order parameters in
different spin states yields the supercurrent of spins. 
In particular, when the system is spontaneously magnetized, 
the spin supercurrent is expected to give a nontrivial effect on the
magnetization dynamics.  
This is the case for a ferromagnetic BEC.
In recent experiments, {\it in situ} techniques for the imaging 
of magnetization profiles enables us to investigate the real-time 
dynamics of magnetizations, such as spin texture formation 
and the nucleation of spin vortices, in ferromagnetic 
BECs~\cite{Sadler2006, berkeley08, Vengalattore2010}. 

For the investigation of the magnetization dynamics, a hydrodynamic
equation has an advantage. It provides the simple description of
magnetization dynamics in a ferromagnetic BEC 
to investigate instabilities~\cite{lama,Kudo} and
configurations of skyrmions and spin textures~\cite{Barnett2009,Cherng2011}.
The hydrodynamic equation, in the absence of energy dissipation,
takes the same form as the Landau-Lifshitz-Gilbert (LLG) equation 
without damping if the partial time derivative is replaced by the
material derivative $D_t=\partial_t+{\bm v}_{\rm mass}\cdot \bm\nabla$, 
or equivalently, if the
adiabatic spin-transfer torque is added~\cite{LLG_ad,LLG_ad2}. 
Here, ${\bm v}_{\rm mass}$ is the superfluid velocity, which is
related to the magnetization direction $\hat{\bm f}$ as
\begin{align}
\nabla\times {\bm v}_{\rm mass} = \frac{\hbar F}{2M}\hat{\bm f}\cdot 
(\nabla\hat{\bm f}\times \nabla\hat{\bm f}),
\label{eq:MH}
\end{align}
where $F$ and $M$ are the spin and mass of an atom, respectively.
This identity comes from the continuous spin-gauge symmetry of the
ferromagnetic BEC~\cite{Ho1996, Nakahara2000}, 
and is known as the Mermin-Ho relation~\cite{Mermin}.
Since the spin is transferred along the velocity field 
${\bm v}_{\rm mass}$ [see Eq.~\eqref{eq:v_spin}], 
the appearance of ${\bm v}_{\rm mass}$ in the hydrodynamic equation 
is the consequence of the spin supercurrent. 

In this paper, we investigate the effect of 
${\bm v}_{\rm mass}$ on the magnetization dynamics using the
hydrodynamic description. 
We show that in the presence of energy dissipation,
the Gross-Pitaevskii (GP) equation for a ferromagnetic BEC is reduced to
a dissipative hydrodynamic equation, which is equivalent to
the extended LLG equation written as
\begin{align}
 \frac{\partial \hat{\bm{f}}}{\partial t}
 = \frac{1}{\hbar}\hat{\bm{f}} \times \bm{B}_{\rm eff}
- \Gamma'\hat{\bm{f}} \times 
\frac{\partial \hat{\bm{f}}}{\partial t}
- (\bm{v}_{\rm mass}\cdot\nabla)\hat{\bm{f}},
\label{eq.extLLG.0}
\end{align}
where $\bm{B}_{\rm eff}$ is an effective magnetic field and
$\Gamma'$ is a damping parameter.
The standard LLG equation, which is widely used to describe the magnetization
dynamics in ferromagnets~\cite{LLGrev}, consists of a spin torque due to the
effective magnetic field and a damping term, and it corresponds to
Eq.~(\ref{eq.extLLG.0}) without the third term on the right hand side. 
The extended LLG equation
was introduced to describe the magnetization dynamics 
affected by spin currents~\cite{Tatara}, which includes additional torque
terms, the so-called adiabatic~\cite{LLG_ad,LLG_ad2} and
nonadiabatic~\cite{LLG_nonad} spin-transfer torques. 
The third term on the right-hand side of Eq.~(\ref{eq.extLLG.0}) 
corresponds to the adiabatic spin torque term.
Thus, Eq.~(\ref{eq.extLLG.0}) is the extended LLG equation without
the nonadiabatic spin-transfer torque.
In conducting ferromagnets, currents can be controlled by the 
external field as well as generated
by magnetic texture dynamics~\cite{Wong}.
In the case of a ferromagnetic BEC, the spin-transfer torques are
related to the superfluid velocity ${\bm v}_{\rm mass}$, which is
induced by spin textures [see Eq.~(\ref{eq:MH})]. 
The damping parameter $\Gamma'$ in Eq.~(\ref{eq.extLLG.0})
corresponds to the so-called Gilbert damping parameter. The Gilbert
damping was introduced originally on a phenomenological basis. 
However, in a conducting ferromagnet system, the damping parameter can
be derived microscopically~\cite{damp_micro}. In the case of a
ferromagnetic BEC, the damping term arises due to collision 
with non-condensed atoms, which is 
introduced in a phenomenological manner. 

The analogy between the dissipative hydrodynamic equation and the
extended LLG equation implies interesting connections between
ferromagnetic BECs and conducting ferromagnets.
For instance, the current-driven motion of domain walls 
and spin vortices, which has been investigated in conducting 
ferromagnets theoretically~\cite{LLG_nonad,LLG_theory,shibata} and 
experimentally~\cite{LLG_exp,Heyne}, can be investigated also in
ferromagnetic BECs by comparison.  
More interesting phenomena such as the anomalous Hall effect,
which is the Hall effect due to the magnetization in a conducting
ferromagnet~\cite{Ye,Onoda03,Onoda04,Taguchi}, 
may be investigated in a ferromagnetic BEC from the viewpoint of 
the interaction between current and spin configuration.
Since there are no impurities in a ferromagnetic BEC, which is also
indicated by the absence of the nonadiabatic term in the hydrodynamic
equation, we can expect to investigate pure adiabatic  
spin-transfer effects in this system.
These interesting connections motivated us to investigate the domain
wall motion and the effect of the superfluid current in a ferromagnetic
BEC. 
In this paper, we demonstrate the magnetic domain pattern dynamics,
which are mainly simulated by the dissipative hydrodynamic equation.

In the study of magnetization dynamics,
we take into account the magnetic dipole-dipole interaction (MDDI) and
the quadratic Zeeman effect. 
The MDDI is known to yield the spatial structure of
magnetizations~\cite{Landau,magnetic_domains,Kawaguchi2006,Yi2006}. 
On the other hand, the quadratic Zeeman energy determines the easy axis
of the magnetization. 
In this paper, we consider a quasi-two dimensional (2D) system and
choose the easy axis normal to the 2D plane. 
Labyrinthine or striped patterns then appear in the magnetic domains
with the magnetization parallel and anti-parallel to the normal
direction, similar to the domain patterns in a ferromagnetic thin
film~\cite{Deutsch}. 
We numerically investigate the domain formation dynamics with and
without the superfluid current, 
and find that the superfluid current helps the spin transport to reach
a stationary configuration. 
We also discuss characteristic lengths of a domain pattern 
in the stationary state. The analytical estimation of domain size shows
good agreement with the averaged domain size of the numerically simulated
domain pattern.

The paper is organized as follows. 
In Sec.~\ref{sec:hydro},
we derive the dissipative hydrodynamic equation from the GP
equation with dissipation.
In Sec.~\ref{sec:MDDI}, we explain how the MDDI is implemented in the
hydrodynamic equation and which type of magnetic domain pattern is
expected to appear depending on the balance between the MDDI energy and the
quadratic Zeeman energy. 
We demonstrate the dynamics of the magnetic domain patterns 
simulated by the dissipative hydrodynamic
equation in Sec.~\ref{sec:pattern}. The time evolution of average
longitudinal magnetization, kinetic and MDDI energies is also shown
for different quadratic Zeeman energies. The domain pattern dynamics are
compared between hydrodynamic equation simulations with and
without the superfluid current and those of the GP equation.
In Sec.~\ref{sec:length}, we theoretically estimate the characteristic
lengths of magnetic domain patterns. The characteristic domain size shortly
after the emergence of a pattern is estimated from the dynamical
instability, and that of the stationary pattern is
estimated using an ansatz of a stripe domain configuration. 
Conclusions and outlook are given in Sec.~\ref{sec:conc}.

\section{\label{sec:hydro} Dissipative hydrodynamic equation}

We consider a spin-$F$ BEC of $N$ atoms under a uniform magnetic
field applied in the $z$ direction confined in a spin-independent
optical trap 
$U_{\rm trap}(\bm{r})$. The zero-temperature mean-field energy is given
by 
\begin{align}
 \mathcal{E} &= \mathcal{E}_{\rm kin} + \mathcal{E}_{\rm trap} 
+ \mathcal{E}_p + \mathcal{E}_q + \mathcal{E}_{\rm s} 
+ \mathcal{E}_{\rm dd},
\end{align}
where $\mathcal{E}_{\rm kin}$, $\mathcal{E}_{\rm trap}$, 
$\mathcal{E}_p$, $\mathcal{E}_q$, 
$\mathcal{E}_{\rm s}$, and $\mathcal{E}_{\rm dd}$ are the
kinetic energy, the trapping potential energy,
the linear and quadratic Zeeman energies, the
short-range interaction energy, and the MDDI energy, respectively. 
The kinetic and the trapping potential energies are
given by 
\begin{align}
 \mathcal{E}_{\rm kin} &= \int d \bm{r} \sum_{m=-F}^F \Psi_m^*(\bm{r})
\left(
-\frac{\hbar^2}{2M}\nabla^2\right) \Psi_m(\bm{r}),
\\
 \mathcal{E}_{\rm trap} &= \int d \bm{r} U_{\rm trap}(\bm{r})
\sum_{m=-F}^F |\Psi_m(\bm{r})|^2,
\end{align}
respectively, 
where $\Psi_m(\bm{r})$ is the condensate wave function for the atoms
in the magnetic sublevel $m$.
The wave function is normalized to satisfy 
\begin{equation}
 N=\int d\bm{r} \sum_{m=-F}^F |\Psi_m(\bm{r})|^2.
\end{equation}

The linear and quadratic Zeeman energies under the external magnetic field
$\bm{B}=B\hat{z}$ are given by
\begin{align}
 \mathcal{E}_p &= p \int d \bm{r} \sum_{m,n=-F}^F 
\Psi^* _m(\bm{r})(F_z)_{mn} \Psi_n(\bm{r}),
\\
 \mathcal{E}_q &= q \int d \bm{r} \sum_{m,n=-F}^F 
\Psi^* _m(\bm{r}) (F_z^2)_{mn} \Psi_n(\bm{r}),
\label{eq.Eq0}
\end{align}
respectively. Here, $F_{x,y,z}$ are the spin-$F$ matrices.
The linear Zeeman energy per atom is given by $p=g_F\mu_{\rm B} B$, where
$g_F$ is the hyperfine $g$-factor, and $\mu_{\rm B}$ is the Bohr magneton.
The quadratic Zeeman energy is induced by a linearly polarized
microwave field as well as by an external magnetic field: 
$q=q_B+q_{\rm EM}$, where $q_B=(g_F\mu_{\rm B}B)^2/E_{\rm hf}$, with
$E_{\rm hf}$ being the hyperfine splitting energy, and 
$q_{\rm EM}=-\hbar^2\Omega^2/(4\delta)$, with $\Omega$ being the Rabi
frequency and $\delta$ the detuning~\cite{Gerbier}.

The short-range interaction energy is
\begin{align}
 \mathcal{E}_{\rm s} &= \frac12 \int d \bm{r} \sum_{m,n}\sum_{m',n'}
\Psi^*_m(\bm{r})\Psi^* _{m'}(\bm{r})
\sum_{S=0,{\rm even}}^{2F}\sum_{M_S=-S}^S \frac{4\pi\hbar^2}{M} a_S
\langle Fm,Fm'|SM_S \rangle \langle SM_S|Fn',Fn \rangle 
\Psi_{n'}(\bm{r})\Psi_n(\bm{r}),
\end{align}
which comes from the
short-range part of the two-body interaction given by 
\begin{align}
V_{\rm s}(\bm r,\bm r') = \delta(\bm r-\bm r')
\sum_{S=0,{\rm even}}^{2F}\frac{4\pi\hbar^2}{M}a_S\mathcal{P}_S, 
\label{eq.Vs}
\end{align}
where $\mathcal{P}_S=\sum_{M_S=-S}^S|S M_S\rangle \langle S M_S|$
projects a pair of spin-$F$ atoms onto the state with total spin $S$, 
$a_S$ is the $s$-wave scattering length for the corresponding spin
channel $S$,
and $\langle Fm,Fn|S M_S\rangle$ in Eq.~(\ref{eq.Vs}) is the Clebsch-Gordan
coefficient. 

The MDDI energy is given by
\begin{align}
  \mathcal{E}_{\rm dd} &= \frac{c_{\rm dd}}{2} \int d \bm{r} d \bm{r}' 
\sum_{\mu,\nu=x,y,z}f_\mu(\bm{r})Q_{\mu\nu}(\bm{r}-\bm{r}')
f_\nu(\bm{r}'),
\end{align}
where $c_{\rm dd} = \mu_0(g_F \mu_{\rm B})^2/(4\pi)$, 
with $\mu_0$ being the magnetic permeability of the vacuum,
$Q_{\mu\nu}(\bm{r})$
is the dipole kernel, whose detailed form is given in the next section,
and 
\begin{align}
f_\mu(\bm{r}) = \sum_{m,n=-F}^F \Psi^*_m(\bm{r})(F_\mu)_{mn}\Psi_n(\bm{r})
\label{eq.f_mu}
\end{align}
is the spin density.
The number density is defined by
\begin{equation}
 n_{\rm tot}(\bm{r}) = \sum_{m=-F}^F |\Psi_m(\bm{r})|^2.
\label{eq.n_tot}
\end{equation}

The hydrodynamic equation without dissipation has been derived from the
GP equation~\cite{lama,Barnett2009,Kudo}. 
Here, we consider dissipation, which can be
phenomenologically introduced to the GP equation 
by replacing $i\partial /\partial t$ with 
$(i-\Gamma)\partial /\partial t$~\cite{Tsubota}. 
The origin of the dissipation can be interpreted as the relaxation
process of the thermal particles into the
condensate~\cite{Tsubota,Choi}.
The value of $\Gamma$ is often taken to be 0.03, and, in fact, 
experimental results have been well explained by the dissipative equation
with $\Gamma=0.03$~\cite{Tsubota,Choi}.
The dissipative GP equation is given by
\begin{align}
 (i-\Gamma)\hbar\frac{\partial}{\partial t}\Psi_m(\bm{r},t) 
&=
\frac{\delta (\mathcal{E}-N\mu(t))}{\delta \Psi^*_m(\bm{r},t)}
\nonumber\\
&= \sum_{n=-F}^F \left[
-\frac{\hbar^2}{2M}\nabla^2\delta_{mn} + H_{mn}(\bm{r},t) 
-\mu(t)\delta_{mn}
\right] \Psi_n(\bm{r},t),
\label{eq.GP}
\end{align} 
where we use the time-dependent chemical potential $\mu(t)$ so that the
total number of atoms is conserved.
The spin-dependent part $H_{mn}$ is given by 
\begin{align}
H_{mn}(\bm{r},t) 
&= U_{\rm trap}(\bm{r})\delta_{mn} + p(F_z)_{mn} + q(F_z^2)_{mn} 
\nonumber \\
&\quad + \sum_{m',n'=-F}^F 
\sum_{S=0,{\rm even}}^{2F}\sum_{M_S=-S}^S \frac{4\pi\hbar^2}{M} a_S
\langle Fm,Fm'|SM_S \rangle \langle SM_S|Fn',Fn \rangle 
\Psi^*_{m'}(\bm{r},t)\Psi_{n'}(\bm{r},t) 
\nonumber \\
&\quad + c_{\rm dd} \sum_{\mu=x,y,z}b_\mu(\bm{r},t)(F_\mu)_{mn},
\label{eq.G_mn}
\end{align}
where the non-local dipole field $\bm{b}(\bm{r},t)$ is defined by
\begin{equation}
 b_\mu(\bm{r},t) =\int d \bm{r}' \sum_{\nu=x,y,z} Q_{\mu\nu}
(\bm{r}-\bm{r}') f_\nu(\bm{r}',t).
\end{equation}
Below we omit the summation symbol: Greek indices that appear twice
are to be summed over $x$, $y$, and $z$, 
and Roman indices are to be summed over $-F,\ldots,F$. 

In the following,
we consider a ferromagnetic BEC.
We assume that the BEC is fully magnetized, $|\bm f|=Fn_{\rm tot}$, and
only the direction of the spin density can vary in space. 
This assumption is valid 
when the ferromagnetic interaction energy is sufficiently large in
comparison with the other
spinor interaction energies, MDDI energy, quadratic Zeeman energy, 
and the kinetic energy arising from 
the spacial variation of the direction of $\bm f$.
The linear Zeeman effect is not necessarily weaker than
the ferromagnetic interaction, since it merely induces the Larmor precession.
For example, the short-range interaction~\eqref{eq.Vs} for a spin-1 BEC
can be written as (see Appendix~\ref{sec:s-int})
\begin{align}
 \langle mn |V_{\rm s}(\bm r, \bm r') |m'n'\rangle 
= \delta(\bm r-\bm r') \left[c_0 \delta_{mn}\delta_{m'n'} 
+ c_1 (F_\mu)_{mn}(F_\mu)_{m'n'}\right],
\label{eq.V_s}
\end{align}
where $c_0=4\pi\hbar^2(2a_2+a_0)/(3M)$ and $c_1=4\pi\hbar^2(a_2-a_0)/(3M)$.
The ground state is ferromagnetic for $c_1<0$.
The above assumption is valid when $|q|\ll |c_1|n_{\rm tot}$,
$c_{\rm dd}\ll |c_1|$, and
the length scale of the spatial spin structure is larger than the spin
healing length $\xi_{\rm sp}=\hbar/\sqrt{2M|c_1|n_{\rm tot}}$. 
Moreover, in the incompressible limit, namely when the spin-independent
interaction ($c_0 n_{\rm tot}$ for the case of a spin-1 BEC) is much
stronger than the ferromagnetic interaction and MDDI, 
the number density $n_{\rm tot}$ is determined, regardless of the spin
structure, and assumed to be stationary.
This is the case for the spin-1 $^{87}$Rb BEC.

We introduce a normalized spinor $\zeta_m$ with
$\Psi_m(\bm{r},t)=\sqrt{n_{\rm tot}(\bm{r},t)}\zeta_m(\bm{r},t)$. 
When the atomic spin is polarized in the $z$ direction, 
the order parameter  is
given by $\zeta_m^{(0)}=\delta_{mF}$. The general order parameter is
obtained by performing the gauge
transformation and Euler rotation as 
\begin{align}
 \zeta_m &= e^{i\phi}e^{-iF_z\alpha}e^{-iF_y\beta}e^{-iF_z\gamma}
\zeta_m^{(0)} \nonumber\\
 &= e^{i(\phi-F\gamma)}e^{-iF_z\alpha}e^{-iF_y\beta}\zeta_m^{(0)}
\nonumber \\
 &\equiv e^{i(\phi-F\gamma)} U \zeta_m^{(0)},
\label{eq.zeta_general}
\end{align}
where $\alpha$, $\beta$, and $\gamma$ are Euler angles and
$\phi$ is the overall phase. 
Due to the spin-gauge symmetry of the ferromagnetic BEC (i.e.,
the equivalence between the phase change $\phi$ and spin rotation $\gamma$),
distinct configurations of $\zeta_m$ are characterized with a set
of parameters $\alpha$, $\beta$, and $\phi'\equiv\phi-F\gamma$. 
The unit vector of the spin density, 
$\hat{\bm f}\equiv\bm{f}/(Fn_{\rm tot})$, for the order
parameter~(\ref{eq.zeta_general}) is denoted by
$\alpha$ and $\beta$:
\begin{align}
 \hat{\bm f} &= \frac{1}{F} \zeta_m^* {\bm F}_{mn} \zeta_n \nonumber\\
 &= \frac{1}{F} \zeta_m^{(0)*} \left( U^\dagger {\bm F} U
\right)_{mn} \zeta_n^{(0)} \nonumber\\
 &= \frac{1}{F} \mathcal{R}
\left[\zeta_m^{(0)*}{\bm F}_{mn} \zeta_n^{(0)}\right] \nonumber\\
 &= \begin{pmatrix} 
\sin\beta\cos\alpha \\ \sin\beta\sin\alpha \\ \cos\beta 
\end{pmatrix},
\end{align}
where $\mathcal{R}$ is an SO(3) rotation matrix given by
\begin{align}
\mathcal{R}=
\begin{pmatrix}
\cos\alpha & -\sin\alpha & 0 \\
\sin\alpha & \cos\alpha & 0 \\
0 & 0 & 1
\end{pmatrix}
\begin{pmatrix}
\cos\beta & 0 & \sin\beta \\
0 & 1 & 0 \\
-\sin\beta & 0 & \cos\beta 
\end{pmatrix}.
\label{eq:R}
\end{align}

The time evolutions of the total number density and the normalized 
spin density are given by
\begin{align}
 \frac{\partial n_{\rm tot}}{\partial t}
&= \Psi_{m}^*\left( \frac{\partial}{\partial t}\Psi_m \right)
+ \left( \frac{\partial}{\partial t}\Psi_m^* \right) \Psi_m,
\label{eq.dndt.0}
\\
 \frac{\partial \hat{\bm{f}}}{\partial t}
&= \frac{1}{Fn_{\rm tot}}\left[
\Psi_m^*\bm{F}_{mn}\left( \frac{\partial}{\partial t}\Psi_n \right)
+ \left( \frac{\partial}{\partial t}\Psi_m^* \right)
\bm{F}_{mn} \Psi_n
\right]
- \frac{\partial n_{\rm tot}}{\partial t}
\frac{\hat{\bm{f}}}{n_{\rm tot}},
\label{eq.dfdt.0}
\end{align}
respectively. The velocities of the superfluid current and spin superfluid
current are defined by 
\begin{align}
 \bm{v}_{\rm mass} 
&= \frac{\hbar}{2Mi} 
[\zeta_m^*(\nabla\zeta_m) - (\nabla\zeta_m^*)\zeta_m]
\nonumber \\
&= \frac{\hbar}{M}[\nabla\phi' - F(\nabla\alpha)\cos\beta],
\label{eq.vmass}
\\
 \bm{v}_{\rm spin}^\mu 
&= \frac{\hbar}{2Mi} (F_\mu)_{mn}
[\zeta_m^*(\nabla\zeta_n) - (\nabla\zeta_m^*)\zeta_n]
\nonumber \\
&= F \bm{v}_{\rm mass}\hat{f}_\mu - \frac{\hbar F}{2M}
(\hat{\bm{f}}\times\nabla\hat{\bm{f}})_\mu, 
\label{eq:v_spin}
\end{align}
respectively.
Substituting Eq.(\ref{eq.GP}) into Eq.~(\ref{eq.dndt.0}), we obtain
\begin{align}
 \frac{\partial n_{\rm tot}}{\partial t}
&= -\frac{1}{1+\Gamma^2}\nabla\cdot(n_{\rm tot}\bm{v}_{\rm mass})
- \frac{\Gamma}{1+\Gamma^2}\frac{2}{\hbar}n_{\rm tot}
[\mu_{\rm local}-\mu(t)],
\label{eq.dndt}
\end{align}
where
\begin{align}
\mu_{\rm local}
&= \frac{\Psi_{m}^*H_{mn}\Psi_n}{n_{\rm tot}}
-\frac{\hbar^2}{2M}\frac{\nabla^2\sqrt{n_{\rm tot}}}{\sqrt{n_{\rm tot}}}
+ \frac{M}{2}\bm{v}_{\rm mass}^2 
+ \frac{\hbar^2F}{4M}(\nabla\hat{\bm{f}})^2.
\label{eq.mu_local}
\end{align}
From the above equations, 
we can also derive the time derivative of $\bm{v}_{\rm mass}$:
\begin{align}
 M\frac{\partial}{\partial t} \bm{v}_{\rm mass} 
&= - \nabla\left[
\frac{\mu_{\rm local} - \mu(t)}{1+\Gamma^2}
- \frac{\Gamma}{1+\Gamma^2} \frac{\hbar}{2n_{\rm tot}}
\nabla\cdot(n_{\rm tot}\bm{v}_{\rm mass})
\right]
+ \hbar F(\nabla\hat{\bm{f}})\cdot\left(
\hat{\bm{f}}\times\frac{\partial\hat{\bm{f}}}{\partial t}
\right).
\label{eq.dvdt.0}
\end{align}
The detailed derivation is given in Appendix~\ref{sec:A}.

Next, we consider the incompressible limit and
assume $\partial n_{\rm tot}/\partial t=0$. 
Then, Eq.~(\ref{eq.dndt}) leads to
\begin{align}
 \nabla\cdot(n_{\rm tot}\bm{v}_{\rm mass}) 
= -\frac{2}{\hbar}\Gamma n_{\rm tot}\left[
\mu_{\rm local} - \mu(t) \right]. 
\label{eq.nv}
\end{align}
This equation simplifies Eq.~(\ref{eq.dvdt.0}):
\begin{align}
 M\frac{\partial}{\partial t} \bm{v}_{\rm mass} 
&= \frac{\hbar}{2n_{\rm tot}\Gamma}\nabla\left[ 
\nabla\cdot(n_{\rm tot}\bm{v}_{\rm mass})  \right]
+ \hbar F(\nabla\hat{\bm{f}})\cdot\left(
\hat{\bm{f}}\times\frac{\partial\hat{\bm{f}}}{\partial t}
\right).
\label{eq.dvdt}
\end{align}
Substituting Eq.(\ref{eq.GP}) into Eq.~(\ref{eq.dfdt.0}) and using 
Eq.~(\ref{eq.nv}), we obtain the equation of motion for spin as follows:
\begin{align}
 \frac{\partial \hat{\bm{f}}}{\partial t}
&= \frac{1}{1+\Gamma^2}\left[
\frac{1}{\hbar}\hat{\bm{f}} \times \bm{B}_{\rm eff}
- (\bm{v}_{\rm mass}\cdot\nabla)\hat{\bm{f}}
\right]
- \frac{\Gamma}{1+\Gamma^2}\hat{\bm{f}} \times\left[
\frac{1}{\hbar}\hat{\bm{f}} \times \bm{B}_{\rm eff}
- (\bm{v}_{\rm mass}\cdot\nabla)\hat{\bm{f}}
\right],
\label{eq.extLLG}
\\
\bm{B}_{\rm eff} &= \frac{\hbar^2}{2M}\nabla^2\hat{\bm{f}}
+ \frac{\hbar^2}{2M}(\bm{a}\cdot\nabla)\hat{\bm{f}}
- c_{\rm dd}\bm{b} - p\hat{z}
- q(2F-1)\hat{f}_z\hat{z},
\label{eq.B_eff}
\end{align}
where $\bm{a}=(\nabla n_{\rm tot})/n_{\rm tot}$.
The detailed derivation is given in Appendix~\ref{sec:B}.
The effective field $\bm{B}_{\rm eff}$ is also derived from the reduced 
Hamiltonian $\mathcal{H}_{\rm mag}$ that contains only spin-dependent
terms:  
$F\bm{B}_{\rm eff}=-\delta\mathcal{H_{\rm mag}}/\delta\bm{\hat{f}}$, where
\begin{align}
 \mathcal{H}_{\rm mag} &= \frac{1}{n_{\rm tot}}\int d\bm{r} \left\{
\frac{\hbar^2}{4M}\frac{(\nabla\bm{f})^2}{Fn_{\rm tot}} 
+ \frac{c_{\rm dd}}{2}\bm{b}\cdot\bm{f}
+ p\hat{z}\cdot\bm{f}
+ \frac{q}{2}\frac{(2F-1)}{Fn_{\rm tot}}(\hat{z}\cdot\bm{f})^2
\right\},
\end{align}
and $\bm{b}$ includes $\bm{f}$.

Here we note that Eq.~(\ref{eq.extLLG}) is equivalent to
Eq.~(\ref{eq.extLLG.0}) and corresponds to the extended LLG equation, 
which describes the magnetization dynamics in a 
conducting ferromagnet in the presence of spin currents interacting with
magnetization, 
with the adiabatic spin-transfer torque. 
The superfluid velocity $\bm{v}_{\rm mass}$ in Eq.~(\ref{eq.extLLG}) 
corresponds to the spin current, which is associated with the electric current
density in the extended LLG equation of a conducting ferromagnet.

\section{\label{sec:MDDI}Magnetic dipole-dipole interaction}

Below, we consider a BEC confined in a quasi-2D trap whose
Thomas-Fermi radius in the $z$ direction is smaller
than the spin healing length. We approximate the wave function in the
$z$ direction by a Gaussian with width $d$: 
$\Psi_m(\bm{r}_\perp, z)=\psi_m(\bm{r}_\perp)h(z)$,
where $\bm{r}_\perp\equiv (x,y)$, and 
$h(z)=\exp[-z^2/(4d^2)]/(2\pi d^2)^{1/4}$. 
When we consider a quasi-2D BEC, $n_{\rm tot}$, $\hat{\bm{f}}$, and 
$\bm{v}_{\rm mass}$ are defined by means of $\psi_m$ instead of
$\Psi_m$~\cite{kawa09}. 
If one replaces $\Psi_m$
with $\psi_m$, $a_S$ with $\eta a_S$, $c_{\rm dd}$ with 
$\eta c_{\rm dd}$, and $\bm{b}$ with $\bar{\bm{b}}$, 
the equation is the same as Eq.~(\ref{eq.GP}), where 
$\eta=\int dzh^4(z)/\int dzh^2(z)=1/\sqrt{4\pi d^2}$ and
\begin{align}
 \bar{b}_\mu(\bm r_\perp) = \int d^2r'_\perp Q_{\mu\nu}^{\rm (2D)}
(\bm r_\perp - \bm r'_\perp) 
\left[\psi_m^*(\bm r'_\perp)(F_\nu)_{mn}\psi_n(\bm r'_\perp)\right],
\label{eq.bbar}
\end{align}
with
\begin{align}
 Q^{\rm (2D)}_{\mu\nu}({\bm r}_\perp-{\bm r}_\perp') 
= \frac{1}{\eta} \iint dz dz'
h^2(z)h^2(z') Q_{\mu\nu}({\bm r}-{\bm r}').
\label{eq.Q_3to2D}
\end{align}

The 2D dipole kernel in the laboratory frame of reference is given by
\begin{align}
 Q^{\rm (2D,lab)}_{\mu\nu}(\bm{r}) &= \sum_{\bm k} 
e^{i\bm k \cdot \bm r}
\tilde{Q}^{\rm (2D,lab)}_{\bm k \mu\nu},
\label{eq.Q.A}
\end{align}
where the subscript $\perp$ is omitted for simplicity and
\begin{align}
 \tilde{Q}^{\rm (2D,lab)}_{\bm{k}}
&= 
- \frac{4\pi}{3} \left(
\begin{array}{ccc}
 1 & 0 & 0 \\
 0 & 1 & 0 \\
 0 & 0 & -2
\end{array}
\right)
+ 4\pi G(kd) \left(
\begin{array}{ccc}
 \hat{k}_x^2 & \hat{k}_x\hat{k}_y & 0 \\
 \hat{k}_x\hat{k}_y & \hat{k}_y^2 & 0 \\
 0 & 0 & -1
\end{array}
\right),
\label{eq.Qk.A}
\end{align}
with $\bm{k} = (k_x,k_y)$, $k = |\bm{k}|$, $\hat{k}_{x,y} = k_{x,y}/k$,
and 
$G(k) \equiv 2 k e^{k^2} \int_k^\infty e^{-t^2}dt = \sqrt{\pi}ke^{k^2}\mathrm{erfc}(k)$.
It can be shown that $G(k)$ is a monotonically increasing function
that satisfies $G(0)=0$ and $G(\infty)=1$.

When the linear Zeeman energy is much larger than the MDDI energy,
we choose the rotating frame of reference in spin space by replacing
$\Psi_m$ with $e^{-ipmt/\hbar}\Psi_m$, 
and eliminate the linear Zeeman term from the GP equation.
In this case, the contribution of the MDDI is time averaged due to the
Larmor precession.
The 2D dipole kernel, which is averaged over the Larmor precession
period, under an external magnetic field in the $z$ direction, 
is given by~\cite{kawa09} 
\begin{align}
 Q^{\rm (2D, rot)}_{\mu\nu}(\bm r)
  &=  \left(\delta_{\mu\nu}-3\delta_{z\mu}\delta_{z\nu}\right)
\sum_{\bm k}e^{i\bm k \cdot \bm r}
\tilde{\mathcal{Q}}_{\bm k},
\label{eq.Q.B}
\end{align}
where
\begin{align}
  \tilde{\mathcal{Q}}_{\bm k}
=\frac{2\pi}{3}[-2 + 3G(kd)].
\label{eq.Qk.B}
\end{align}

The formation of a stable magnetic domain pattern depends on the
quadratic energy as well as the MDDI.  
Since the quadratic Zeeman energy is the monotonic function of
$q(\hat{z}\cdot\bm{f})^2$, transverse ($\hat{f}_z=0$) and 
longitudinal ($\hat{f}_z=\pm 1$) magnetization is
preferable for $q>0$ and $q<0$, respectively.
For $q>0$, the uniform pattern of transverse magnetization is stable
because of the MDDI.
When a strong magnetic field is applied, the dipole kernel is given by
Eq.~(\ref{eq.Q.B}), which is 
low for small-$k$ modes of a transverse magnetization pattern and for
large-$k$ modes of a longitudinal one.
Thus, a uniform transverse
magnetization pattern (i.e., $k=0$) is stable for $q>0$. 
The situation is the
same for a magnetic pattern under zero field, in which the dipole
kernel is given by Eq.~(\ref{eq.Q.A}). 
For $q<0$, the uniform transverse magnetization pattern is still
stable if the quadratic Zeeman energy is smaller than the MDDI energy.
When $q<0$ and $|q|$ is large enough to balance with the MDDI,
the uniform transverse magnetization becomes unstable and
a longitudinal magnetization pattern appears. 
In other words, there is a threshold of $q$ where a non-uniform
longitudinal magnetization pattern appears. 
The threshold can be calculated from the linear stability analysis,
as will be discussed in Sec.~\ref{sec:length}.

Here, we focus on the negative-$q$ regime, which can be achieved 
by means of a linearly polarized microwave field
even in the absence of an external magnetic field 
[see, below Eq.~(\ref{eq.Eq0})].
When $q<0$ and $|q|$ is sufficiently large, longitudinal
magnetization is dominant and magnetic domains with 
$\hat{f}_z\simeq \pm 1$ form patterns because of the MDDI.
This situation is consistent with that of a uni-axial 
ferromagnet~\cite{Deutsch}.

\section{\label{sec:pattern} Domain formation dynamics}

In this section, 
we focus on the domain formation dynamics with and without 
$\bm{v}_{\rm mass}$ under a strong magnetic field
to illustrate how $\bm{v}_{\rm mass}$ affects domain
pattern formation. 
Later, the domain patterns 
under zero magnetic field are also shown as
complementary results. 
The validity of the dissipative hydrodynamic equation is
also examined by comparing hydrodynamic 
and GP equation simulations.

For the magnetic domain pattern simulations, 
we solve the coupled equations (\ref{eq.dvdt}) and
(\ref{eq.extLLG}), where ${\bm v}_{\rm mass}$
and ${\bm B}_{\rm eff}$ are defined by Eqs. (\ref{eq.vmass}),
and (\ref{eq.B_eff}), respectively.
For simplicity, we take $\bm{a}=0$ and use periodic boundary conditions. 
In the case of a strong magnetic field, we employ 
Eq.~(\ref{eq.Q.B}) as a dipole kernel. 
To see the role of $\bm{v}_{\rm mass}$, 
we demonstrate the calculation without $\bm{v}_{\rm mass}$, in which
$\bm{v}_{\rm mass}$ is always taken to be zero, as well as 
the full calculation using all those equations in the
presence of $\bm{v}_{\rm mass}$. 
The initial condition is $\hat{f}_x\simeq 1$, $\hat{f}_y\simeq 0$, and
$\hat{f}_z\simeq 0$ with small noises, and $\bm{v}_{\rm mass}=0$.

\begin{figure}[tb]
\includegraphics[width=4.5cm,clip]{hydro_q50_t.eps}
\includegraphics[width=3cm,clip]{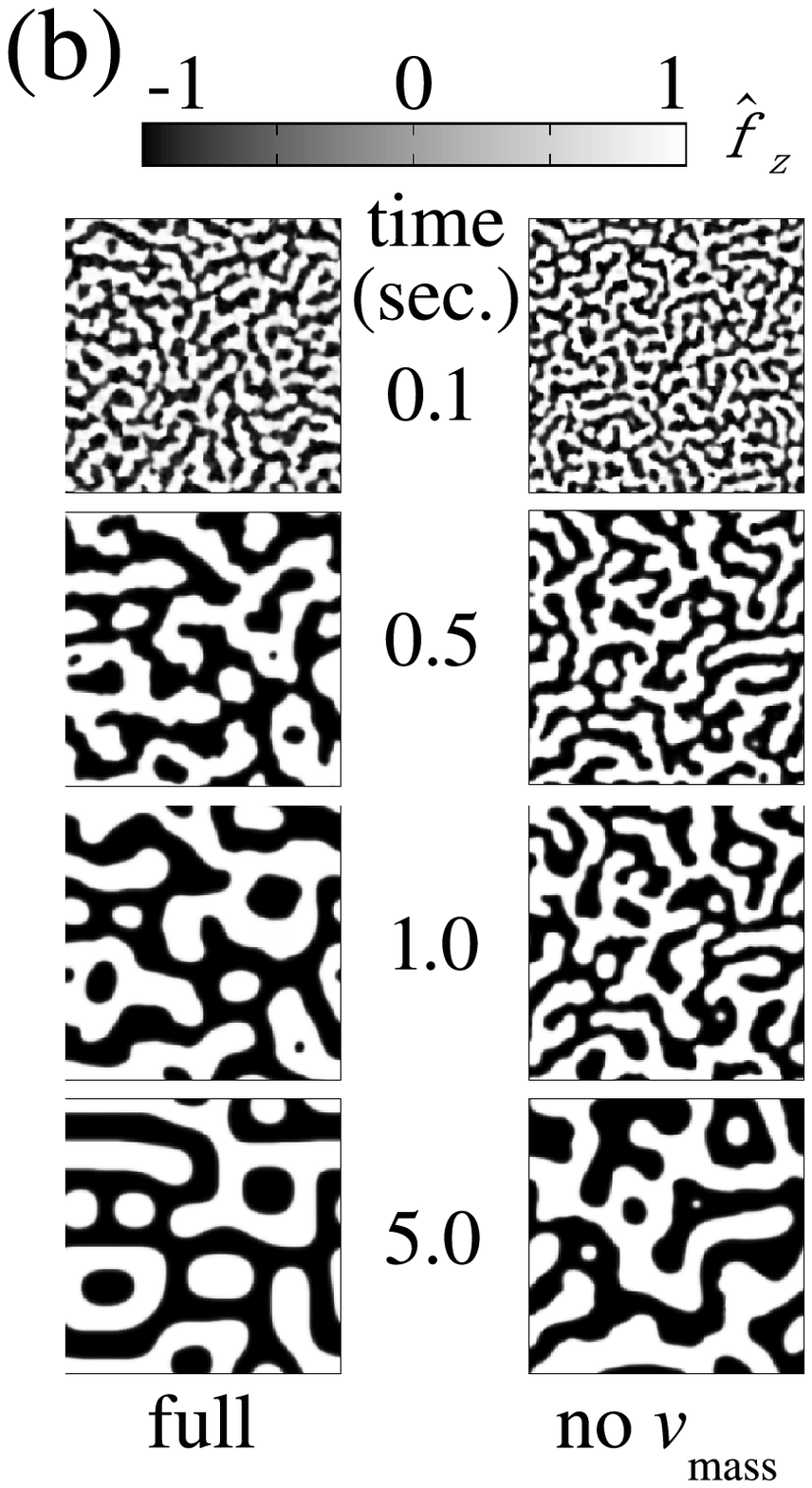}
\caption{ (Color online) (a) Time dependence of the averaged
 longitudinal magnetization, kinetic and MDDI energies in
 the presence (solid curves) and absence (dashed curves) of 
 $\bm{v}_{\rm mass}$. The dot-dashed curve in the graph of $E_{\rm kin}/h$
(middle panel)
 corresponds to the contribution from $\bm{v}_{\rm mass}$.
(b) Snapshots of longitudinal magnetization in which white and black
 correspond to positive and negative values of $\hat{f}_z$,
 respectively. The size of each snapshot is 
$256 \times 256$ $\mu$m.
Here, we take $q/h=-50$ Hz, and 
$n_{\rm tot}= 6n_{\rm tot}^{(0)}$, where 
$n_{\rm tot}^{(0)}=\sqrt{2\pi d^2}n_{\rm 3D}^{(0)}$ with  
$n_{\rm 3D}^{(0)}=2.3\times 10^{14}$ cm$^{-3}$ and $d=1.0$ $\mu$m. 
The damping rate is given by a typical value $\Gamma=0.03$.
The other parameters are given by the typical values for a spin-1
$^{87}$Rb atom: $M = 1.44 \times 10^{-25}$ Kg, $F=1$, and $g_F = -1/2$.}
\label{fig:hydro_q50} 
\end{figure}

Figure~\ref{fig:hydro_q50}(a) shows that the longitudinal magnetization
grows rapidly to form magnetic domains. 
The kinetic and MDDI energies also grow rapidly at first.
After the rapid increase, the kinetic energy decays as magnetic domains
grow and the domain wall density decreases. 
The averaged longitudinal magnetization, kinetic energy and its
contribution from $\bm{v}_{\rm mass}$, and MDDI energy 
(per unit area per atom) are defined by
\begin{align}
 \overline{|\hat{f}_z|} &= \frac{1}{L^2}\int d^2r \; |\hat{f}_z(\bm{r})|,
\label{eq.fbar}
\\
 E_{\rm kin} &= E_{\rm flow} + \frac{1}{L^2}\int d^2r \;
  \frac{\hbar^2}{4M}\left[
(\nabla\hat{f}_x(\bm{r}))^2 + (\nabla\hat{f}_y(\bm{r}))^2 
+ (\nabla\hat{f}_z(\bm{r}))^2 
\right],
\label{eq.E_kin}
\\
 E_{\rm flow} &= \frac{1}{L^2}\int d^2r \;
\frac{M}{2}\bm{v}_{\rm mass}(\bm{r})^2 ,
\label{eq.E_flow}
\\
 E_{\rm dd} &= \frac{1}{L^2}\int d^2r \;
\frac{c_{\rm dd}}{2}\eta\bar{\bm{b}}(\bm{r})\cdot\hat{\bm{f}}(\bm{r}),
\label{eq.E_dd}
\end{align}
respectively. Here, $L$ is the system size.

The longitudinal magnetization grows rapidly for a short
time, and then the averaged longitudinal magnetization saturates 
at around $\overline{|\hat{f}_z|}\simeq 1$ [see the top panel of
Fig.~\ref{fig:hydro_q50}(a)]. 
Figure~\ref{fig:hydro_q50}(b) shows that 
magnetic domains emerge in a short time and, after that,  
spread slowly to reach a stationary configuration.  
After the rapid growth of $\hat{f}_z$, magnetic domain patterns grow
faster in the presence than in the absence of $\bm{v}_{\rm mass}$. 
The difference is apparent in the series of snapshots and $E_{\rm kin}$
and $E_{\rm dd}$, although the contribution of $\bm{v}_{\rm mass}$ to 
$E_{\rm kin}$ is small compared with the total kinetic energy [see the
middle panel of Fig.~\ref{fig:hydro_q50}(a)]. 
In the presence of $\bm{v}_{\rm mass}$, magnetic domains grow
efficiently because of spin transfer accompanied by the transfer of
atoms by means of $\bm{v}_{\rm mass}$.

\begin{figure}[tb]
\includegraphics[width=7cm,clip]{hydro_GP_t.eps}
\\
\includegraphics[width=6cm,clip]{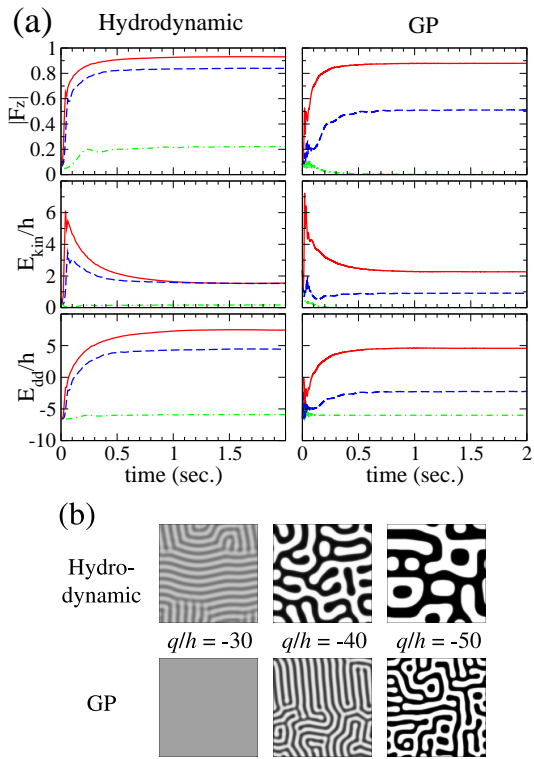}
\caption{(Color online) Comparison of domain formation dynamics
 between the hydrodynamic description and GP numerics.
(a) Time dependence of the averaged
 longitudinal magnetization, kinetic and MDDI energies simulated by 
the hydrodynamic equation (left column) and 
the GP equation (right column). 
(b) Snapshots at $t=5$ s simulated by the hydrodynamic and GP equations.
Solid, dashed, and dot-dashed curves in (a)
are for $q/h=-50$, $-40$, and $-30$, respectively.}
\label{fig:hydro_GP} 
\end{figure}

To examine the validity of the hydrodynamic description, we compare the
simulations of the dissipative hydrodynamic and the GP
equations.
In both cases, the averaged $|\hat{f}_z|$, $E_{\rm kin}$, 
$E_{\rm dd}$, and snapshots for several values of $q$ 
are shown in Fig.~\ref{fig:hydro_GP}.
For the GP equation, the averaged longitudinal magnetization,
kinetic and MDDI energies (per unit area per atom) are defined by
\begin{align}
 \overline{|\hat{f}_z|} &= \frac{1}{L^2}\int d^2r \; 
\frac{|f_z(\bm{r})|}{n_{\rm tot}(\bm{r})},
\label{eq.fbar.GP}
\\
 E_{\rm kin} &= \frac{1}{L^2}\frac{1}{N}\left(
-\frac{\hbar^2}{2M}\right)\int d^2r \sum_m
\psi_m^*(\bm{r})\nabla^2\psi_m(\bm{r}),
\label{eq.E_kin.GP}
\\
 E_{\rm dd} &= \frac{1}{L^2}\frac{1}{N}\int d^2r \;
\frac{c_{\rm dd}}{2}\eta\bar{\bm{b}}(\bm{r})\cdot\bm{f}(\bm{r}),
\label{eq.E_dd.GP}
\end{align}
respectively. Here ${\bm f}$ and $n_{\rm tot}$ are defined in 
Eqs.~(\ref{eq.f_mu}) and (\ref{eq.n_tot}),
respectively.
These values are equal to Eqs. (\ref{eq.fbar}), (\ref{eq.E_kin}), 
and (\ref{eq.E_dd}) if $n_{\rm tot}$ is
uniform and $|\bm{f}(\bm{r})|=n_{\rm tot}$. 
Below the threshold where a non-uniform pattern begins to appear, 
the most stable pattern is a uniform pattern. 
The threshold is $q/h\simeq -30$ for the hydrodynamic
description and $q/h\simeq -40$ for the GP equation.  
Near the threshold,
the averaged longitudinal magnetization is small compared to those in 
other cases. 
Except for in the vicinity of the threshold, the time dependence of domain
pattern formation looks similar for the two simulations.
However, the domain size differs significantly due to differences in the
domain wall structure. 
In the GP simulation, the amplitude of magnetization is
suppressed in domain walls, while the suppression of magnetization
does not occur 
in the hydrodynamic description. 
An analysis
of domain sizes for both cases will be given in the next section.

\begin{figure}[tb]
\includegraphics[width=7cm,clip]{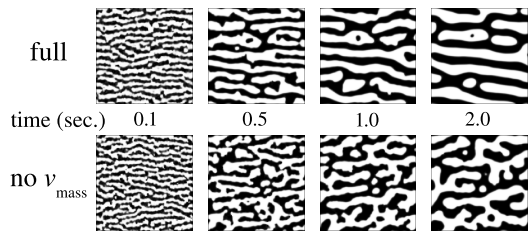}
\caption{Domain pattern formation in the absence of a
 magnetic field. The direction of the stripes depends on the initial
 configuration of spins. The parameters are the same as those given in
 Fig.~\ref{fig:hydro_q50}}.  
\label{fig:B0_q50} 
\end{figure}

The magnetic domain patterns under zero field 
are simulated by the dissipative hydrodynamic equations with
Eq.~(\ref{eq.Qk.A}). 
In Fig.~\ref{fig:B0_q50}, domain pattern formation from the
initial condition of $\hat{f}_x\simeq 1$ is shown in the
presence and in the absence of $\bm{v}_{\rm mass}$.
The magnetic domain patterns under zero field look similar to those
under a strong magnetic field. The effect of $\bm{v}_{\rm mass}$ is also
similar: $\bm{v}_{\rm mass}$ moves the domain walls faster.
However, they are strongly affected by the initial condition. 
This is because the spin and orbit degrees of freedom couple in the
MDDI under zero field [see Eq.~(\ref{eq.Qk.A})].
If the initial condition is given by $\hat{f}_y\simeq 1$, 
one can see remarkably similar domain patterns as shown in the
snapshots in Fig.~\ref{fig:B0_q50}, which are rotated by 90$^\circ$.

\section{\label{sec:length} Characteristic lengths of domain patterns}

We have seen in the previous section that 
the magnetic domain pattern has an initially short characteristic length
that later increases in size. 
At the beginning of domain pattern formation,
domain size is estimated from the linear instability analysis. 
The domain size of the stationary pattern
can be estimated as that of a stripe domain pattern.
For the parameters given in our numerical simulations, the
domain size of the stable stripe pattern is longer than the domain size
estimated from the linear instability.

\subsection{Dynamical instability}

The dynamical instability (linear instability) under a strong magnetic
field has been discussed previously, both for the hydrodynamic
equation~\cite{Kudo} and for the GP equation~\cite{kawa09}.
For the hydrodynamic equation, the growth
rate of the unstable mode is calculated from the eigenvalues of the
linearized equation of small deviations from the uniform initial 
condition. For the GP equation, a similar equation is
derived by means of Bogoliubov analysis.
When the initial magnetic pattern is uniform and $\hat{f}_x=1$, 
the respective growth rates of the unstable mode for the hydrodynamic and GP
equations are given by
\begin{align}
 \lambda^{\rm H}(k)&= \sqrt{ \left[
-\frac{\hbar^2k^2}{2M} - q - 4\pi c_{\rm dd}\tilde{n}_{\rm tot}[1-G(kd)]
\right] \left[
\frac{\hbar^2k^2}{2M} + 2\pi c_{\rm dd}\tilde{n}_{\rm tot}G(kd)
\right]},
\label{eq.lamH}
\\
 \lambda^{\rm G}(k)&= \sqrt{ \left[
-\frac{\hbar^2k^2}{2M} - q - 4\pi c_{\rm dd}\tilde{n}_{\rm tot}
(1-\tilde{q})[1-G(kd)]
\right] \left[
\frac{\hbar^2k^2}{2M} + 2\pi c_{\rm dd}\tilde{n}_{\rm tot}(1+\tilde{q})G(kd)
\right]},
\label{eq.lamG}
\end{align}
where $\tilde{n}_{\rm tot}=n_{\rm tot}\eta$ and 
\begin{align}
 \tilde{q} &= \frac{q}{2\tilde{n}_{\rm tot}(|c_1|+4\pi c_{\rm dd}/3)}.
\end{align}
Here, we have considered $\bm{v}_{\rm mass}=0$ and neglected the
dissipation, which is very small (i.e., $\Gamma\ll 1$).

The domain size at the emergence of a non-uniform pattern
is estimated as
\begin{equation}
 \ell_{\rm i} = \pi/k_0,
\label{eq.l_i}
\end{equation}
where $k_0$ is given by the momentum at which 
$\lambda^{\rm H}(\bm{k}_0)$ or $\lambda^{\rm G}(\bm{k}_0)$ has its
maximum value.

\subsection{Domain size estimated from the hydrodynamic equation}

Here, we estimate characteristic lengths of the domain pattern in the
stationary state.
For $q<0$ with large $|q|$, the
ideal stable pattern is the stripe pattern of longitudinal
magnetization. We assume that the stable pattern is described by 
\begin{align}
 \hat{f}_x = \mathrm{cn}(x/\kappa\xi,\kappa^2), \quad
\hat{f}_y = 0, \quad
 \hat{f}_z = \mathrm{sn}(x/\kappa\xi,\kappa^2),
\end{align}
where $\mathrm{sn}(x/\kappa\xi,\kappa^2)$ and 
$\mathrm{cn}(x/\kappa\xi,\kappa^2)$ are the Jacobi elliptic
functions with $0<\kappa^2\le 1$. 
These functions contain the characteristic lengths of the stripe domain
pattern~\cite{Ezawa}: the domain wall width $\xi$ 
and the periodicity of the pattern $2\ell_{\rm s}$ with
\begin{align}
 \ell_{\rm s} \equiv 2\kappa\xi K(\kappa^2),
\end{align}
where $K(\kappa^2)$ is the complete elliptic integral of the first kind. 
The kinetic and quadratic Zeeman energies are calculated as 
\begin{align}
 E_{\rm kin} &= \frac{\hbar^2}{2M}\frac{1}{\xi\ell_{\rm s}}
\frac{E(\kappa^2)}{\kappa},
\\
 E_q &= \frac{1}{L^2}\int d^2 r \; \frac{q}{2}(1+\hat{f}_z^2)
\nonumber \\
 &= \frac{q}{2} -\frac{q\xi}{\ell_{\rm s}}\frac{E(\kappa^2)}{\kappa},
\end{align}
where $E(\kappa^2)$ is the complete elliptic integral of the second kind. 
The domain wall width is estimated to be the length at which the
summation of the two energies has a minimum value. Solving 
$\partial (E_{\rm kin} + E_q)/\partial \xi=0$, we obtain 
\begin{equation}
 \xi = \frac{\hbar}{\sqrt{-2Mq}}.
\label{eq.xi_HD}
\end{equation} 

For estimation of $\ell_{\rm s}$, we need to take the MDDI energy into
account. For simplicity, we assume $\ell_{\rm s}\gg\xi$, and take 
$\hat{f}_x=\hat{f}_y=0$ and 
\begin{align}
 \hat{f}_z &= \left\{
\begin{array}{cc}
 -1 & \mbox{for}\quad (2n-1)\ell_{\rm s} < x < 2n\ell_{\rm s}\\
  1 & \mbox{for}\quad  2n\ell_{\rm s} < x < (2n+1)\ell_{\rm s}
\end{array}
\right.
\nonumber \\
&= \sum_{n=1}^\infty \frac{1-(-1)^n}{in\pi}
(e^{in\pi x/\ell_{\rm s}}-e^{-in\pi x/\ell_{\rm s}}),
\end{align}
where $n$ is an integer. From Eqs.~(\ref{eq.bbar}), (\ref{eq.Q.B}), and
(\ref{eq.E_dd}), we can calculate the MDDI energy as   
\begin{align}
 E_{\rm dd}= - \frac{4\pi}{3}c_{\rm dd}\tilde{n}_{\rm tot}
\sum_{n=1}^\infty \left[
-2 + 3G(n\pi d/\ell_{\rm s})
\right] \left[
\frac{1-(-1)^n}{n\pi}
\right]^2.
\label{eq.E_dd.l0}
\end{align}
Instead of using the original definition of $G(k)$, we use an
approximate function $G(k)\simeq 1-\exp(-\sqrt{\pi}k)$. Then,
Eq.~(\ref{eq.E_dd.l0}) is rewritten as
\begin{align}
 E_{\rm dd}= - \frac{4\pi}{3}c_{\rm dd}\tilde{n}_{\rm tot}\left\{
\frac12 + \frac{6}{\pi^2}\left[ 
\mathrm{Li}_2( -e^{-\pi\sqrt{\pi} d/\ell_{\rm s}} )
- \mathrm{Li}_2( e^{-\pi\sqrt{\pi} d/\ell_{\rm s}} )
\right]\right\},
\label{eq.E_dd.HD}
\end{align}
where $\mathrm{Li}_s(z)\equiv \sum_{k=0}^\infty z^k/k^s$ is the
polylogarithm. We estimate $\ell_{\rm s}$ to be the length at which the
total energy has its minimum value 
in the limit of $\kappa\to 1$ [thus $E(\kappa^2)\to 1$].
Solving $\partial (E_{\rm kin}+E_q+E_{\rm dd})/\partial \ell_{\rm s}=0$ and 
taking $\kappa=1$, we obtain
\begin{align}
 \ell_{\rm s} = \frac{\sqrt{\pi}^3 d}{2}\frac{1}{\mathrm{arctanh}\left[
\exp\left(-\frac{\hbar^2/(M\xi)}{8c_{\rm dd}\tilde{n}_{\rm tot}\sqrt{\pi}d}
\right)\right]}.
\label{eq.ell.HD}
\end{align}

\begin{figure}[tb]
\includegraphics[width=5cm,clip]{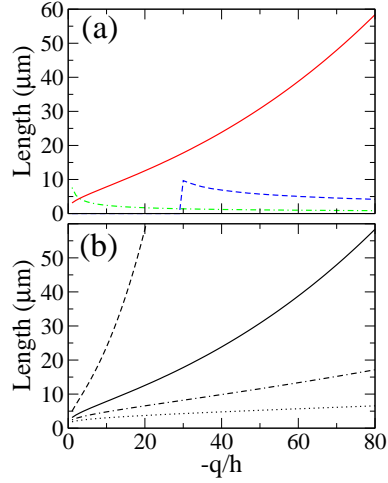}
\caption{(Color online) Dependence of theoretical characteristic
 lengths on $-q/h$ for the hydrodynamic description. 
(a) Theoretically estimated characteristic lengths
 of a stripe pattern (solid curve) [Eq.~(\ref{eq.ell.HD})], 
a pattern at its emergence (dashed curve) 
[Eq.~(\ref{eq.l_i}) with Eq.~(\ref{eq.lamH})], 
and the domain wall width (dot-dashed curve) [Eq.~(\ref{eq.xi_HD})]
for $n_{\rm tot}=6n_{\rm tot}^{(0)}$. 
(b) Theoretical characteristic length
 of a stripe pattern [Eq.~(\ref{eq.ell.HD})]
for $n_{\rm tot}=3n_{\rm tot}^{(0)}$ (dashed 
 line), $6n_{\rm tot}^{(0)}$ (solid line), $10n_{\rm tot}^{(0)}$ (dot-dashed
 line), and $20n_{\rm tot}^{(0)}$ (dotted line).} 
\label{fig:length_HD} 
\end{figure}

In  Fig.~\ref{fig:length_HD}(a), 
we plot the $q$ dependence of $\ell_{\rm i}, \xi$, and $\ell_{\rm s}$ 
for $n_{\rm tot}=6n_{\rm tot}^{(0)}$. 
The estimated domain size is in good agreement with the average domain
size of the simulations; e.g., for $q/h=-40$, 
the estimated domain size $\ell_{\rm s}\simeq 24$ $\mu$m from 
Eq.~(\ref{eq.ell.HD}) and the average domain size
$\overline{\ell_{\rm s}}\sim 20$ $\mu$m from simulations shown 
in Fig.~\ref{fig:hydro_GP}(b). 
The stable domain size $\ell_{\rm s}$ is larger 
than the initial domain size $\ell_{\rm i}$. 
This property is consistent with the simulations in which
initially small magnetic domains appear within a short
time before spreading to form a stable (or metastable) pattern. 
The $q$ dependence of $\ell_{\rm s}$ is also consistent with the
simulations in which the domain size increases as $|q|$ increases. 

The domain size of a stable pattern depends
also on the number density. In Fig.~\ref{fig:length_HD}(b), 
$\ell_{\rm s}$ is plotted for $n_{\rm tot}=3n_{\rm tot}^{(0)}$, 
$6n_{\rm tot}^{(0)}$, $10n_{\rm tot}^{(0)}$ and $20n_{\rm tot}^{(0)}$.
The smaller the value of $n_{\rm tot}$, the larger the $\ell_{\rm s}$. 
For the typical number density $n_{\rm tot}=n_{\rm tot}^{(0)}$ in
experiments~\cite{Sadler2006,berkeley08,Vengalattore2010}, 
the estimated domain size is too large to observe in a conventional
experimental system.

\subsection{Domain size estimated from the GP equation}

In the GP equation, spins are not always fully magnetized. 
Even when $\hat{f_z}\simeq \pm 1$ over most of a magnetic domain pattern, 
the magnetization in the domain walls may vanish.
The domain wall structure depends on the ferromagnetic 
interaction ($c_1 n_{\rm tot}$), which competes with the quadratic Zeeman
and kinetic energies. 
The ferromagnetic interaction lowers energy for high spin density. 
In other words, the transverse magnetization exists in the domain wall
between magnetic domains with $\hat{f}_z\simeq\pm 1$,
when the ferromagnetic interaction is strong.
On the other hand, the quadratic Zeeman energy prefers a state with
sublevels $m=\pm 1$.
Here, let us consider two states with $\hat{f}_z=0$; for instance, (a) 
$(\zeta_1,\zeta_0,\zeta_{-1})^{T}=(1/2,1/\sqrt{2},1/2)^{T}$ 
and (b) $(1/\sqrt{2},0,1/\sqrt{2})^{T}$.
State (a) corresponds to $\hat{f}_x=1$ and 
$\hat{f}_y=\hat{f}_z=0$ (transverse magnetization), and
state (b) to  $\hat{f}_x=\hat{f}_y=\hat{f}_z=0$ (zero magnetization).
Comparing them, one can find that the quadratic Zeeman energy of state
(b) is smaller than that of state (a).
In other words, domain walls with zero magnetization appear when the
quadratic Zeeman energy is dominant.
Therefore, the domain wall structure changes at around 
$|c_1|\tilde{n}_{\rm tot} \simeq |q|$.

\begin{figure}[tb]
\includegraphics[width=5cm,clip]{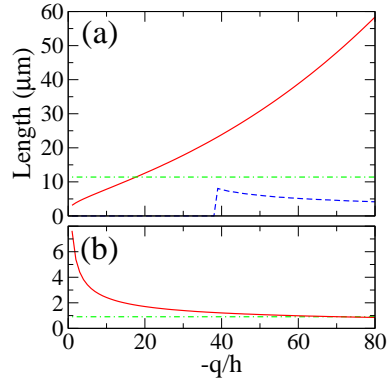}
\caption{(Color online) Dependence of theoretical characteristic
 lengths on $-q/h$ for the GP equation. 
 (a) Theoretically estimated characteristic lengths
 of the stripe pattern with fully-magnetized domain walls 
(solid curve) [Eq.~(\ref{eq.ell.HD})], 
that of the stripe pattern without transverse magnetization 
(dot-dashed line) [Eq.~(\ref{eq.ell.GP})], 
and that of a pattern at its emergence (dashed curve) 
[Eq.~(\ref{eq.l_i}) with Eq.~(\ref{eq.lamG})] for 
$n_{\rm tot}=6n_{\rm tot}^{(0)}$. 
(b) Theoretically estimated domain width of the stripe pattern with
 fully-magnetized domain walls (solid curve) [Eq.~(\ref{eq.xi_HD})] and 
that of the stripe pattern without transverse magnetization 
(dot-dashed line) [Eq.~(\ref{eq.xi_GP})] for  
$n_{\rm tot}=6n_{\rm tot}^{(0)}$.} 
\label{fig:length_GP} 
\end{figure}

Now we estimate characteristic lengths of magnetic domains for 
$|c_1|\tilde{n}_{\rm tot} < |q|$. Since zero magnetization is
preferred in domain walls, we assume the stripe pattern in this
case is described by
\begin{align}
 \psi_1=\sqrt{n_{\rm tot}}\cos\frac{\theta-\pi/2}{2}, \quad
 \psi_0=0, \quad
 \psi_{-1}=\sqrt{n_{\rm tot}}\sin\frac{\theta-\pi/2}{2},
\label{eq.psi_GP}
\end{align}
with $\theta=\mathrm{am}(x/\kappa\xi,\kappa^2)$, where 
$\mathrm{am}(x/\kappa\xi,\kappa^2)$ is the Jacobi amplitude.
Equation~(\ref{eq.psi_GP}) corresponds to
\begin{align}
 \hat{f}_x=\hat{f}_y=0, \quad 
\hat{f}_z=\mathrm{sn}(x/\kappa\xi,\kappa^2).
\end{align}
Then, the kinetic, quadratic Zeeman, and ferromagnetic interaction
energies are calculated as  
\begin{align}
 E_{\rm kin} &= \frac{\hbar^2}{4M}\frac{1}{\xi\ell_{\rm s}}
\frac{E(\kappa^2)}{\kappa},
\\
 E_q &= q,
\\
 E_{c1} &= \frac{c_1\tilde{n}_{\rm tot}}{2}\frac{1}{L^2}\int dr^2
\; \hat{\bm{f}}^2(\bm{r})
\nonumber \\
&= \frac{c_1\tilde{n}_{\rm tot}}{2}\left[
\frac{1}{\kappa^2} - \frac{2\xi}{\ell_{\rm s}}
\frac{E(\kappa^2)}{\kappa}
\right].
\end{align}
Note that $E_q$ is independent of $\xi$ and $\ell_{\rm s}$. 
Solving $\partial(E_{\rm kin} + E_{c1})/\partial\xi=0$ in the 
$\kappa\to 1$ limit, we obtain
\begin{align}
 \xi = \frac{\hbar}{\sqrt{-4Mc_1\tilde{n}_{\rm tot}}}.
\label{eq.xi_GP}
\end{align}
In Fig.~\ref{fig:length_GP}(b), Eq.~(\ref{eq.xi_GP}) is plotted
as the dot-dashed line, while the solid curve is the plot of
Eq.~(\ref{eq.xi_HD}). Equation~(\ref{eq.xi_GP}) is valid for 
$|q|>|c_1|\tilde{n}_{\rm tot}\simeq 35h$.
In this region, the values of $\xi$ of both equations almost coincide, 
as seen in Fig.~\ref{fig:length_GP}(b).  

To estimate the domain size of the stripe pattern, 
we again assume the MDDI energy is given by Eq.~(\ref{eq.E_dd.HD}). 
Solving $\partial(E_{\rm kin} + E_{c1} + E_{\rm dd})/\partial\ell_{\rm s}=0$ 
with $\kappa=1$, we obtain
\begin{align}
 \ell_{\rm s} = \frac{\sqrt{\pi}^3 d}{2}\frac{1}{\mathrm{arctanh}\left[
\exp\left(-\frac{\hbar^2/(M\xi)}{16c_{\rm dd}\tilde{n}_{\rm tot}\sqrt{\pi}d}
\right)\right]}.
\label{eq.ell.GP}
\end{align}
In Fig.~\ref{fig:length_GP}(a), Eq.~(\ref{eq.ell.GP}) is plotted
as the dot-dashed line and compared with the solid curve given by
Eq.~(\ref{eq.ell.HD}). 
The dashed curve expresses the domain size $\ell_{\rm i}$ at the
beginning of domain pattern formation and has a nonzero value above the
threshold. Since $|q|>|c_1|\tilde{n}_{\rm tot}$ in the region above the
threshold, the domain size $\ell_{\rm s}$ is estimated by
Eq.~(\ref{eq.ell.GP}) instead of Eq.~(\ref{eq.ell.HD}) for 
the stable pattern simulated by the GP equation.
Also in this case, the estimated domain size is in good agreement with 
the average domain size obtained from simulations. 

\section{\label{sec:conc} Conclusions and outlook}

We have derived the dissipative hydrodynamic equation of a ferromagnetic
Bose-Einstein condensate (BEC). This equation has the same form as the
extended Landau-Lifshitz-Gilbert (LLG) equation, which was originally 
developed to explain the spin dynamics in a conducting 
ferromagnet interacting with spin-polarized currents, 
including an adiabatic spin-transfer torque term. 
The dissipative hydrodynamic equation enables us to investigate 
how the domain formation dynamics are affected by the superfluid velocity
$\bm{v}_{\rm mass}$, which is inseparable in the Gross-Pitaevskii (GP)
equation. 

We have demonstrated domain pattern formation simulated by the
dissipative hydrodynamic equation with and without $\bm{v}_{\rm mass}$.
Although no remarkable difference appears at the beginning of domain
pattern formation, $\bm{v}_{\rm mass}$ has an effect on later domain
formation dynamics: pattern formation is faster in the
presence than in the absence of $\bm{v}_{\rm mass}$.
We have also shown simulations of the GP equation and
compared them with those of the hydrodynamic equation. 
The dependence on $q$, which characterizes the quadratic Zeeman energy,
of domain pattern formation
is different between hydrodynamic and GP simulations because
the threshold for a nonuniform magnetic pattern differs
between them.
Nevertheless, for large $|q|$, magnetic domain patterns eventually come
to look similar in both simulations, 
although they differ in size.
The difference is caused by the fact that the assumption of full
magnetization for the hydrodynamic description is not satisfied for
large $|q|$ in the GP simulations.

To explain the difference in domain size
between the hydrodynamic and GP equation simulations, 
we have estimated the characteristic lengths of the domain patterns. 
The domain size at the beginning of pattern formation is estimated
by means of the linear stability analysis. The difference in domain
size for a short time is based on the dynamical instability.
The domain size of the domain pattern in the stationary configuration 
is estimated using the ansatz of the stripe pattern of longitudinal
magnetization domains. 
The analytical estimations are in good agreement with 
the numerical simulations. 

In conclusion, the dissipative hydrodynamic equation provides a simple
approach to discuss magnetization dynamics in a ferromagnetic BEC.
The hydrodynamic equation simulation qualitatively well reproduces
the domain formation dynamics that are simulated by the GP equation.
However, quantitative discrepancies arise when the assumption of the
hydrodynamic description fails: for instance, the ferromagnetic
interaction energy becomes comparable with the quadratic Zeeman energy. 
The analogy between the dissipative hydrodynamic equation and 
the extended LLG equation can provide suggestions on new experiments of
a ferromagnetic BEC to investigate interesting phenomena that are
observed in conducting ferromagnets. 

One of such interesting and possible phenomena is the anomalous Hall
effect (AHE), which is the Hall 
effect due to the magnetization and observed in conducting ferromagnets. 
Here, we focus on the AHE caused by a skyrmion configuration 
or spin chirality~\cite{Ye,Onoda03,Onoda04,Taguchi}, 
although there are several mechanisms to cause the phenomenon.  
The Berry phase, which is generated by a skyrmion configuration, induces 
an effective magnetic field or gauge flux. When an electric field is
applied, electrons move to the perpendicular direction to both the
electric and the effective magnetic field. 
This is the mechanism of the AHE due to spin chirality.
Inversely, the electric current can move the skyrmion.
The AHE is expected to be observed very clearly 
in the adiabatic limit, where a ferromagnetic BEC is supposed to
realize. Unfortunately, there are difficulties for observation of the AHE
in a ferromagnetic BEC: for example, it is difficult to create the
external field that corresponds to the electric field of a conducting
ferromagnet system. 
However, the essential aspect of the AHE (i.e., the interaction
between current and spin chirality) can be investigated in a
ferromagnetic BEC.
Experimentally, one can create a current by sudden change of a trapping
potential or 
the fictitious field that is created through the vector potential induced
by a laser field~\cite{Lin}.
The investigation about the interaction between current and spin
configuration in a ferromagnetic BEC will give an insight on pure 
adiabatic spin-transfer effects. 

\begin{acknowledgments}
The authors thank M. Ueda for his useful comments. 
This work is supported
by MEXT JSPS KAKENHI (No. 22103005, 22340114, 22740265), the Photon Frontier
Network Program of MEXT, Japan, Hayashi Memorial Foundation for Female Natural
Scientists, and JSPS and FRST under the Japan-New
 Zealand Research Cooperative Program.  
\end{acknowledgments}

\appendix

\section{\label{sec:A} Time evolution of $\bm{v}_{\rm mass}$}

Substituting Eq.~(\ref{eq.zeta_general}) into Eq.~(\ref{eq.GP}) and
applying $\Psi_m^*$ from the left, we have
\begin{align}
 (i-\Gamma)\hbar
&\left( 
\sqrt{n_{\rm tot}}\frac{\partial\sqrt{n_{\rm tot}}}{\partial t}
+ in_{\rm tot}\frac{\partial\phi'}{\partial t}
- in_{\rm tot}F\cos\beta\frac{\partial\alpha}{\partial t}
\right) 
\nonumber\\
&=
-\frac{\hbar^2}{2M}\Psi_m^{(0)*}\left\{
[\nabla + i\nabla\phi' 
- i(\nabla\alpha) e^{iF_y\beta} F_z e^{-iF_y\beta}
- i(\nabla\beta)F_y]^2 \right\}_{mn} \Psi_n^{(0)} 
\nonumber \\
&\quad + \Psi_m^*H_{mn}\Psi_n - n_{\rm tot}\mu(t),
\label{eq.GP.mod}
\end{align}
where we have used $\Psi_m=e^{i\phi'}{U}_{mn}\Psi_n^{(0)}$.
The first term on the right hand side is calculated as 
\begin{align}
 \Psi_m^{(0)*}&\left\{
[\nabla + i\nabla\phi' 
- i(\nabla\alpha) e^{iF_y\beta} F_z e^{-iF_y\beta}
- i(\nabla\beta)F_y]^2 \right\}_{mn} \Psi_n^{(0)} 
\nonumber\\
&= n_{\rm tot}\zeta_m^{(0)*}\left\{
\frac{\nabla^2\sqrt{n_{\rm tot}}}{\sqrt{n_{\rm tot}}}
- [\nabla\phi' - F_z(\nabla\alpha)\cos\beta]^2
- [F_x(\nabla\alpha)\sin\beta - F_y(\nabla\beta)]^2
\right\}_{mn} \zeta_n^{(0)}
\nonumber\\
&\quad
+i\zeta_m^{(0)*}\left\{
n_{\rm tot}\nabla\cdot [\nabla\phi' - F_z(\nabla\alpha)\cos\beta]
+ 2\sqrt{n_{\rm tot}}\nabla\sqrt{n_{\rm tot}}\cdot
[\nabla\phi' - F_z(\nabla\alpha)\cos\beta]
\right\}_{mn} \zeta_n^{(0)}
\nonumber\\
&= n_{\rm tot}\left[
\frac{\nabla^2\sqrt{n_{\rm tot}}}{\sqrt{n_{\rm tot}}}
- \left( \frac{M}{\hbar} \bm{v}_{\rm mass} \right)^2
- \frac{F}{2}(\nabla\hat{\bm{f}})^2
\right]
+ i\frac{M}{\hbar}\nabla\cdot(n_{\rm tot}\bm{v}_{\rm mass}),
\end{align}
where we have used Eq.~(\ref{eq.vmass}) and the following equations:
\begin{align}
\zeta_m^{(0)*} (F_\mu)_{mn} \zeta_n^{(0)}&=F\delta_{\mu z},
\label{eq.a0}
\\ 
e^{iF_y\beta} F_z e^{-iF_y\beta} 
&= F_z\cos\beta  - F_x\sin\beta ,
\label{eq.a1}
\\
\zeta_m^{(0)*} \left\{ \left[
F_x(\nabla\alpha)\sin\beta - F_y(\nabla\beta)  
\right]^2 \right\}_{mn} \zeta_n^{(0)}
&= \frac{F}{2}\left[
(\nabla\alpha)^2\sin^2\beta + (\nabla\beta)^2
\right] 
\nonumber \\
&= \frac{F}{2}(\nabla\hat{\bm{f}})^2 .
\label{eq.a2}
\end{align}

The imaginary part of Eq.~(\ref{eq.GP.mod}) reads
\begin{align}
 \frac{\partial n_{\rm tot}}{\partial t} 
&= 2\Gamma n_{\rm tot}\left(
\frac{\partial\phi'}{\partial t} 
- F\cos\beta\frac{\partial\alpha}{\partial t}
\right)
- \nabla\cdot(n_{\rm tot}\bm{v}_{\rm mass}).
\label{eq.Im}
\end{align}
From the real part of Eq.~(\ref{eq.GP.mod}), we have
\begin{align}
 \frac{\hbar\Gamma}{2n_{\rm tot}}\frac{\partial n_{\rm tot}}{\partial t}
+ \hbar\left(
\frac{\partial\phi'}{\partial t} 
- F\cos\beta\frac{\partial\alpha}{\partial t}
\right)
+ [\mu_{\rm local}-\mu(t)] &= 0,
\label{eq.Re}
\end{align}
where
\begin{align}
 \mu_{\rm local}(\bm{r},t) 
&= \frac{\Psi_m^*H_{mn}\Psi_n}{n_{\rm tot}}
-\frac{\hbar^2}{2M}\frac{\nabla^2\sqrt{n_{\rm tot}}}{\sqrt{n_{\rm tot}}}
+ \frac{M}{2}\bm{v}_{\rm mass}^2 
+ \frac{\hbar^2F}{4M}(\nabla\hat{\bm{f}})^2.
\end{align}
Combining Eqs.~(\ref{eq.Im}) and (\ref{eq.Re}) and eliminating
$(\partial \phi'/\partial t -F\cos\beta\partial \alpha/\partial t)$,  
we obtain Eq.~(\ref{eq.dndt}) as follows:
\begin{align}
 \frac{\partial n_{\rm tot}}{\partial t}
&= -\frac{1}{1+\Gamma^2}\nabla\cdot(n_{\rm tot}\bm{v}_{\rm mass})
- \frac{\Gamma}{1+\Gamma^2}\frac{2}{\hbar}n_{\rm tot}
[\mu_{\rm local}-\mu(t)].
\label{a.dndt}
\end{align}

On the other hand,
substituting Eq.~(\ref{eq.Im}) into Eq.~(\ref{a.dndt}), we have
\begin{align}
 \hbar\left( \frac{\partial \phi'}{\partial t}
- F\cos\beta\frac{\partial\alpha}{\partial t} \right)
&= -\frac{1}{1+\Gamma^2} [\mu_{\rm local} - \mu(t)]
+ \frac{\Gamma}{1+\Gamma^2} \frac{\hbar}{2n_{\rm tot}}
\nabla\cdot(n_{\rm tot}\bm{v}_{\rm mass}).
\label{eq.a4}
\end{align}
The gradient of the left hand side of Eq.~(\ref{eq.a4}) is
\begin{align}
  \hbar\nabla \left( \frac{\partial \phi'}{\partial t}
- F\cos\beta\frac{\partial\alpha}{\partial t} \right)
&= \hbar\left\{
\frac{\partial}{\partial t}[\nabla\phi' - F(\nabla\alpha)\cos\beta]
- F\left[
(\nabla\cos\beta)\frac{\partial\alpha}{\partial t}
- (\nabla\alpha)\frac{\partial\cos\beta}{\partial t}
\right] \right\}
\nonumber \\
&= M\frac{\partial}{\partial t}\bm{v}_{\rm mass}
- \hbar F\left[
(\nabla\alpha)\sin\beta\frac{\partial\beta}{\partial t}
- (\nabla\beta)\sin\beta\frac{\partial\alpha}{\partial t}
\right].
\label{eq.a5}
\end{align}
Here, we introduce vectors $\hat{\bm{m}}$ and $\hat{\bm{n}}$, which are
orthogonal to $\hat{\bm{f}}$,
\begin{align}
 \hat{\bm{f}} = 
\begin{pmatrix}
 \sin\beta\cos\alpha \\ \sin\beta\sin\alpha \\ \cos\beta
\end{pmatrix}
, \quad
 \hat{\bm{m}} = 
\begin{pmatrix}
 \cos\beta\cos\alpha \\ \cos\beta\sin\alpha \\ -\sin\beta
\end{pmatrix}
, \quad
 \hat{\bm{n}} = 
\begin{pmatrix}
 -\sin\alpha \\ \cos\alpha \\ 0
\end{pmatrix}.
\label{eq.vectors}
\end{align} 
One can easily see
\begin{align}
 \hat{\bm{m}} = \hat{\bm{n}} \times \hat{\bm{f}}, \quad
\hat{\bm{n}} = \hat{\bm{f}} \times \hat{\bm{m}}
\label{eq.vecs1}
\\
 \partial\hat{\bm{f}} = (\partial\beta)\hat{\bm{m}} 
+ (\partial\alpha)\sin\beta\hat{\bm{n}}. 
\label{eq.vecs2}
\end{align}
Employing these vectors, we have
\begin{align}
 (\nabla\hat{\bm{f}})\cdot\left(
\hat{\bm{f}} \times \frac{\partial\hat{\bm{f}}}{\partial t}
\right) 
&= [(\nabla\beta)\hat{\bm{m}} + (\nabla\alpha)\sin\beta\hat{\bm{n}}]
\cdot\left[
\hat{\bm{f}}\times\left(
\frac{\partial\beta}{\partial t}\hat{\bm{m}}
+ \frac{\partial\alpha}{\partial t}\sin\beta\hat{\bm{n}}
\right)\right]
\nonumber \\
&= [(\nabla\beta)\hat{\bm{m}} + (\nabla\alpha)\sin\beta\hat{\bm{n}}]
\cdot\left(
\frac{\partial\beta}{\partial t}\hat{\bm{n}}
- \frac{\partial\alpha}{\partial t}\sin\beta\hat{\bm{m}}
\right)
\nonumber \\
&=
(\nabla\alpha)\sin\beta\frac{\partial\beta}{\partial t}
- (\nabla\beta)\sin\beta\frac{\partial\alpha}{\partial t}.
\label{eq.a9}
\end{align}
From Eqs.(\ref{eq.a4}), (\ref{eq.a5}) and (\ref{eq.a9}), 
we obtain Eq.~(\ref{eq.dvdt.0}) as follows:
\begin{align}
 M\frac{\partial}{\partial t} \bm{v}_{\rm mass} 
&= - \nabla\left[
\frac{\mu_{\rm local} - \mu(t)}{1+\Gamma^2}
- \frac{\Gamma}{1+\Gamma^2} \frac{\hbar}{2n_{\rm tot}}
\nabla\cdot(n_{\rm tot}\bm{v}_{\rm mass})
\right]
+ \hbar F(\nabla\hat{\bm{f}})\cdot\left(
\hat{\bm{f}}\times\frac{\partial\hat{\bm{f}}}{\partial t}
\right).
\end{align}

\section{\label{sec:B} Time evolution of $\bm{f}$}

The time evolution of the normalized spin density is calculated as 
\begin{align}
 \frac{\partial\hat{f}_{\mu}}{\partial t}
&= \frac{1}{Fn_{\rm tot}}\left[
\Psi_m^*(F_\mu)_{mn}\frac{\partial \Psi_n}{\partial t} 
+ \frac{\partial \Psi_m^*}{\partial t}(F_\mu)_{mn}\Psi_{n}
\right]
- \frac{\partial n_{\rm tot}}{\partial t}
\frac{\hat{f}_\mu}{n_{\rm tot}}
\nonumber \\
&= \frac{1}{Fn_{\rm tot}}\frac{1}{\hbar(1+\Gamma^2)}\left\{
-\frac{\hbar^2}{2M} \left[
(-i-\Gamma)\Psi_m^*(F_\mu)_{mn}\nabla^2\Psi_n
+ (i-\Gamma)(\nabla^2\Psi_m^*)(F_\mu)_{mn}\Psi_n
\right]
\right.
\nonumber \\
&\quad\quad 
\left.
+ (-i-\Gamma)
\Psi_m^*(F_\mu)_{ml}[H-\mu(t)]_{ln}\Psi_n
+ (i-\Gamma)
\Psi_m^*[H-\mu(t)]_{ml}(F_\mu)_{ln}\Psi_n
\right\}
\nonumber \\
&= \frac{1}{Fn_{\rm tot}}\frac{1}{\hbar(1+\Gamma^2)}\left\{
- \frac{\hbar^2}{M}\left[ \mathrm{Im}\left( 
\Psi_m^*(F_\mu)_{mn}\nabla^2\Psi_n \right)  
- \Gamma \mathrm{Re} \left(
\Psi_m^*(F_\mu)_{mn}\nabla^2\Psi_n \right)\right]
\right.
\nonumber \\
&\quad\quad
\left.
+ 2\left[ \mathrm{Im}\left(
\Psi_m^*(F_\mu)_{ml}H_{ln}\Psi_n \right)
-\Gamma \mathrm{Re} \left(
\Psi_m^*(F_\mu)_{ml}H_{ln}\Psi_n \right)\right]
+ 2\Gamma\mu(t) \right\},
\label{eq.dfdt.1}
\end{align}
where we have assumed $\partial n_{\rm tot}/\partial t=0$.

\subsection{Kinetic energy terms}

First, we calculate the spacial derivative terms.
Making use of Eqs.~(\ref{eq.a1}) and (\ref{eq.vectors}),
we have
\begin{align}
 \Psi_m^*(F_\mu)_{mn}\nabla^2\Psi_n
&= \zeta_m^{(0)*}
(F_x\hat{m}_\mu + F_y\hat{n}_\mu + F_z\hat{f}_\mu)\left\{
\sqrt{n_{\rm tot}}\nabla^2\sqrt{n_{\rm tot}}
\right.
\nonumber\\
&\quad\quad
+ in_{\rm tot}\nabla\cdot\left[
\nabla\phi' - (\nabla\alpha)F_z\cos\beta
+ (\nabla\alpha)F_x\sin\beta - (\nabla\beta)F_y \right]
\nonumber\\
&\quad\quad
+ i\nabla n_{\rm tot}\cdot\left[
\nabla\phi' - (\nabla\alpha)F_z\cos\beta
+ (\nabla\alpha)F_x\sin\beta - (\nabla\beta)F_y \right]
\nonumber\\
&\quad\quad
\left.
- n_{\rm tot}\left[
\nabla\phi' - (\nabla\alpha)F_z\cos\beta
+ (\nabla\alpha)F_x\sin\beta - (\nabla\beta)F_y \right]^2
\right\}_{mn} \zeta_n^{(0)}.
\end{align}
Using Eqs.~(\ref{eq.vmass}), (\ref{eq.a2}),
(\ref{eq.vecs1}), (\ref{eq.vecs2}) and the relation
\begin{align}
 (\nabla\alpha)\sin\beta\hat{\bm{m}} - (\nabla\beta)\hat{\bm{n}}
&= - \hat{\bm{f}}\times\nabla\hat{\bm{f}},
\\
 (\nabla\alpha)\sin\beta\hat{\bm{n}} + (\nabla\beta)\hat{\bm{m}}
&= - \hat{\bm{f}}\times(\hat{\bm{f}}\times\nabla\hat{\bm{f}}),
\end{align}
we obtain
\begin{align}
 \frac{\Psi_m^*(F_\mu)_{mn}\nabla^2\Psi_n}{Fn_{\rm tot}} 
&=- \left\{ -\frac{\nabla^2\sqrt{n_{\rm tot}}}{\sqrt{n_{\rm tot}}}
+ \left( \frac{M}{\hbar} \bm{v}_{\rm mass} \right)^2
+ \frac{F}{2}(\nabla\hat{\bm{f}})^2
\right\}\hat{f}_\mu
+ \frac{M}{\hbar}\left[ \hat{\bm{f}}\times
(\bm{v}_{\rm mass}\cdot\nabla)\hat{\bm{f}}
\right]_\mu
\nonumber\\
&\quad
- \frac12\left[ \hat{\bm{f}}\times\left(
\hat{\bm{f}} \times \nabla^2\hat{\bm{f}}
\right) \right]_\mu
- \frac12\left[ \hat{\bm{f}}\times\left(
\hat{\bm{f}} \times \frac{\nabla n_{\rm tot}}{n_{\rm tot}} 
\cdot\nabla\hat{\bm{f}}
\right) \right]_\mu
\nonumber\\
&\quad
+ i\frac{M}{\hbar}\left[
\frac{\nabla\cdot(n_{\rm tot}\bm{v}_{\rm mass})}{n_{\rm tot}}\hat{f}_\mu
+ (\bm{v}_{\rm mass}\cdot\nabla)\hat{f}_\mu
\right]
\nonumber\\
&\quad
- \frac{i}{2}\left[ 
\hat{\bm{f}} \times \nabla^2\hat{\bm{f}}
\right]_\mu
- \frac{i}{2}\left[ 
\hat{\bm{f}} \times \frac{\nabla n_{\rm tot}}{n_{\rm tot}} 
\cdot\nabla\hat{\bm{f}}
\right]_\mu.
\end{align}
From this equation and Eq.~(\ref{eq.nv}), we can rewrite the space 
derivative terms of Eq.~(\ref{eq.dfdt.1}) as
\begin{align}
 \mathrm{Re}\left[
\frac{\Psi_m^*(F_\mu)_{mn} \nabla^2 \Psi_n}{Fn_{\rm tot}}
\right]
&= \frac{2M}{\hbar^2} \left(
\frac{\Psi_m^*H_{mn}\Psi_n}{n_{\rm tot}} - \mu_{\rm local}
\right)\hat{f}_\mu
\nonumber\\
&\quad
+ \left\{ \hat{\bm{f}}\times \left[ 
\frac{M}{\hbar}(\bm{v}_{\rm mass}\cdot\nabla)\hat{\bm{f}}
- \frac12\hat{\bm{f}}\times\left( \nabla^2\hat{\bm{f}}
+ (\bm{a}\cdot\nabla)\hat{\bm{f}} \right)
\right] \right\}_\mu,
\label{eq.Re_nabla2}
\\
\mathrm{Im}\left[
\frac{\Psi_m^*(F_\mu)_{mn} \nabla^2 \Psi_n}{Fn_{\rm tot}}
\right]
&=
- \Gamma \frac{2M}{\hbar^2}[\mu_{\rm local} - \mu(t)]\hat{f}_\mu
+ \left[ \frac{M}{\hbar} (\bm{v}_{\rm mass}\cdot\nabla)\hat{\bm{f}}
-\frac12\hat{\bm{f}}\times\left( \nabla^2\hat{\bm{f}}
+ (\bm{a}\cdot\nabla)\hat{\bm{f}} \right)
\right]_\mu,
\label{eq.Im_nabla2}
\end{align}
where $\bm{a}=\nabla n_{\rm tot}/n_{\rm tot}$.

\subsection{\label{sec:s-int}Short-range interaction terms}

Next, we calculate the contributions of the short-range interaction 
$\Psi_m^* (F_\mu)_{ml}H^{\rm s}_{ln}\Psi_n$,
where
\begin{align}
H^{\rm s}_{mn} &=  C^{mm'}_{nn'}\Psi^*_{m'}\Psi_{n'}, 
\label{eq:G^s0}\\
 C^{mm'}_{nn'} &\equiv \sum_{S=0}\frac{4\pi\hbar^2}{M} a_S 
\langle Fm,Fm'|\mathcal{P}_S|Fn',Fn\rangle.
\end{align}
Now we rewrite the matrix elements of the short-range interaction 
$C^{mm'}_{nn'}$
in terms of the spin matrices.
The projection operator $\mathcal{P}_S$ satisfies the completeness 
relation $\sum_S \mathcal{P}_S=1$, that is,
\begin{align}
\delta_{mn}\delta_{m'n'} = \sum_{S=0,{\rm even}}^{2F} 
\langle Fm,Fm'|\mathcal{P}_S|Fn',Fn\rangle .
\label{eq:identity1}
\end{align}
On the other hand, from the identity equation
\begin{align}
 (\bm F_1 \cdot \bm F_2)^k &= \left[
\frac{(\bm F_1+\bm F_2)^2-\bm F_1^2 -\bm F_2^2}{2}\right]^k,
\end{align}
we obtain
\begin{align}
&(F_{\nu_1}F_{\nu_2}\cdots F_{\nu_k})_{mn} 
(F_{\nu_1}F_{\nu_2}\cdots F_{\nu_k})_{m'n'} \nonumber\\ 
&=\sum_{S=0,{\rm even}}^{2F}
\left[\frac{S(S+1)-2F(F+1)}{2}\right]^k 
\langle Fm,Fm'|\mathcal{P}_S|Fn',Fn\rangle.
\label{eq:identity2}
\end{align}
Using Eqs.~\eqref{eq:identity1} and \eqref{eq:identity2}, 
$C^{mm'}_{nn'}$ can be generally expressed as
\begin{align}
&C^{mm'}_{nn'}=
 \Lambda_0 \delta_{mn}\delta_{m'n'} 
+\sum_{k=1}^{F} \Lambda_k 
(F_{\nu_1}F_{\nu_2}\cdots F_{\nu_k})_{mn} 
(F_{\nu_1}F_{\nu_2}\cdots F_{\nu_k})_{m'n'},
\label{eq:Cmn}
\end{align}
where $\Lambda_0$ and $\Lambda_k$ are given by the linear 
combinations of $a_S$.
For the case of $F=1$, for example, we obtain
$\Lambda_0=c_0=\pi\hbar^2 (2a_2+a_0)/(3M)$ 
and $\Lambda_1=c_1=4\pi\hbar^2(a_0-a_2)/(3M)$, and
$C^{mm'}_{nn'}$ can be written in the following form:
\begin{align}
 C^{mm'}_{nn'} = c_0\delta_{mn}\delta_{m'n'}
+ c_1(\bm F)_{mn}\cdot(\bm F)_{m'n'}.
\end{align}

Substituting Eq.~\eqref{eq:Cmn} to Eq.~\eqref{eq:G^s0}, we obtain
\begin{align}
H^{\rm s}_{mn} &=  C^{mm'}_{nn'}\Psi^*_{m'}\Psi_{n'}\\
&= \Lambda_0 n_{\rm tot} \delta_{mn} 
+ \sum_{k=1}^{F}  \Lambda_k n_{\rm tot} 
 \mathcal{M}_{\nu_1\nu_2\cdots\nu_k}
(F_{\nu_1}F_{\nu_2}\cdots F_{\nu_k})_{mn} ,
\label{eq:G^s}
\end{align}
where 
\begin{align}
\mathcal{M}_{\nu_1\nu_2\cdots\nu_k}&= \zeta_m^* 
(F_{\nu_1}F_{\nu_2}\cdots F_{\nu_k})_{mn}\zeta_n,
\end{align}
and we have used $\Psi_m=\sqrt{n_{\rm tot}}\zeta_m$ and 
$\zeta_m^* \zeta_m=1$.

When we consider the ferromagnetic state (i.e., 
when the order parameter is given by 
$\Psi_m = \sqrt{n_{\rm tot}}e^{i\phi'}U_{mn}\zeta_n^{(0)}$ 
with $\zeta_m^{(0)}=\delta_{mF}$), 
we have
\begin{align}
 \mathcal{M}_{\nu_1\nu_2\cdots\nu_k} 
&= \zeta_m^{(0)*} (U^\dagger F_{\nu_1} U U^\dagger F_{\nu_2} U 
\cdots U^\dagger F_{\nu_k} U)_{mn}\zeta_n^{(0)} \nonumber\\
&= \mathcal{R}_{\nu_1\nu_1'}\mathcal{R}_{\nu_2\nu_2'}\cdots 
\mathcal{R}_{\nu_k\nu_k'}(F_{\nu_1'} F_{\nu_2'} \cdots 
F_{\nu_k'})_{FF} \nonumber\\
&= \mathcal{R}_{\nu_1\nu_1'}\mathcal{R}_{\nu_2\nu_2'}\cdots 
\mathcal{R}_{\nu_k\nu_k'}\mathcal{M}_{\nu_1'\nu_2'\cdots\nu_k'}^{(0)}, 
\end{align}
where $\mathcal{M}_{\nu_1\nu_2\cdots\nu_k}^{(0)}\equiv (F_{\nu_1} F_{\nu_2} \cdots F_{\nu_k})_{FF}$, 
and $\mathcal{R}$ is defined in Eq.(\ref{eq:R}).
Then, $\Psi_m^* H_{mn}^{\rm s}\Psi_n$ is shown to be independent 
of the local spin direction:
\begin{align}
 \Psi_m^* H_{mn}^{\rm s}\Psi_n = n_{\rm tot}^2\Lambda_0   
+ n_{\rm tot}^2 \sum_{k=1}^{F}\Lambda_k 
\mathcal{M}_{\nu_1\nu_2\cdots\nu_k}^{(0)} 
\mathcal{M}_{\nu_1\nu_2\cdots\nu_k}^{(0)}.
\end{align}
In a similar manner, we obtain
\begin{align}
&\Psi_m^* (F_\mu H^{\rm s})_{mn}\Psi_n
= n_{\rm tot}^2 \Lambda_0 F\hat{f}_\mu 
 + n_{\rm tot}^2 \sum_{k=1}^{F}\Lambda_k \mathcal{R}_{\mu\mu'}
\mathcal{M}_{\nu_1\nu_2\cdots\nu_k}^{(0)} 
\mathcal{M}_{\mu'\nu_1\nu_2\cdots\nu_k}^{(0)}.
\label{eq:FH}
\end{align}
Note here that $F_x$ or $F_y$ has to appear an even number of times 
in the product of $F_{\nu_1} F_{\nu_2} \cdots F_{\nu_k}$ so that
$\mathcal{M}_{\nu_1\nu_2\cdots\nu_k}^{(0)} = (F_{\nu_1} F_{\nu_2} \cdots F_{\nu_k})_{FF}$ 
is nonzero. Hence, 
$\mathcal{M}_{\nu_1\nu_2\cdots\nu_k}^{(0)} \mathcal{M}_{\mu'\nu_1\nu_2\cdots\nu_k}^{(0)}$ 
becomes nonzero only when $\mu'=z$:
\begin{align}
\mathcal{M}_{\nu_1\nu_2\cdots\nu_k}^{(0)} 
\mathcal{M}_{\mu'\nu_1\nu_2\cdots\nu_k}^{(0)} 
= \delta_{\mu'z} F \mathcal{M}_{\nu_1\nu_2\cdots\nu_k}^{(0)}
\mathcal{M}_{\nu_1\nu_2\cdots\nu_k}^{(0)}.
\end{align}
Then, Eq.~\eqref{eq:FH} is written as
\begin{align}
&\Psi_m^* (F_\mu H^{\rm s})_{mn}\Psi_n\nonumber\\
&= n_{\rm tot}^2 \Lambda_0 F\hat{f}_\mu 
 + n_{\rm tot}^2 F \mathcal{R}_{\mu z}\sum_{k=1}^{F}\Lambda_k 
\mathcal{M}_{\nu_1\nu_2\cdots\nu_k}^{(0)} 
\mathcal{M}_{\nu_1\nu_2\cdots\nu_k}^{(0)}\\
&= (\Psi_m^*H^{\rm s}_{mn}\Psi_n) F\hat{f}_\mu.
\end{align}
Thus, we obtain
\begin{align}
{\rm Re}[\Psi_m^* (F_\mu H^{\rm s})_{mn}\Psi_n] 
&=(\Psi_m^*H^{\rm s}_{mn}\Psi_n) F\hat{f}_\mu,\label{eq.Re0}\\
{\rm Im}[\Psi_m^* (F_\mu H^{\rm s})_{mn}\Psi_n] &=0\label{eq.Im0}.
\end{align}

\subsection{Other terms}

Finally, we calculate the remaining terms. 
Using Eqs.~(\ref{eq.G_mn}) and (\ref{eq.Im0}), we easily obtain
\begin{align}
\\
 \mathrm{Im}\left[
\frac{\Psi_m^*(F_\mu)_{ml} H_{ln} \Psi_n}{Fn_{\rm tot}}
\right]
&= \frac{1}{2iFn_{\rm tot}}\Psi_m^*[F_\mu, H ]_{mn}\Psi_n
\nonumber \\
&=\frac{1}{2iFn_{\rm tot}}\Psi_m^*\left\{
p[F_\mu,F_z] + q[F_\mu,F_z^2] + c_{\rm dd}b_\nu[F_\mu,F_\nu]
\right\}_{mn}\Psi_n
\nonumber \\
&= -\frac12 \left[ \hat{\bm{f}} \times \left(
p\hat{z} + q(2F-1)\hat{f}_z\hat{z} + c_{\rm dd}\bm{b} 
\right) \right]_\mu .
\label{eq.Im_FH}
\end{align}
Employing Eqs.~(\ref{eq.G_mn}), (\ref{eq.Re0}), and (\ref{eq.Im0}), 
we obtain
\begin{align}
 \mathrm{Re}\left[
\frac{\Psi_m^*(F_\mu)_{ml} H_{ln} \Psi_n}{Fn_{\rm tot}}
\right]
&= \frac{\Psi_m^*H_{mn}^{\rm s}\Psi_n}{n_{\rm tot}} \hat{f}_\mu
+ U_{\rm trap}\hat{f}_\mu
+ \frac{p}{2}\left[\delta_{\mu z} + (2F-1)\hat{f}_z\hat{f}_\mu \right]
\nonumber \\
&\quad
+ \frac{q}{2}\left[(2F^2-3F+1)\hat{f}_z^2\hat{f}_\mu + F\hat{f}_\mu
+ (2F-1)\hat{f}_z\delta_{\mu z} \right]
\nonumber\\
&\quad
+ \frac{c_{\rm dd}}{2}\left[ b_\mu 
+ (2F-1)(\bm{b}\cdot\hat{\bm{f}})\hat{f}_\mu \right],
\label{eq.b3}
\end{align}
where we have used
\begin{align}
\Psi_m^* (F_\mu F_\nu +  F_\nu F_\mu)_{mn} \Psi_n
&= Fn_{\rm tot}\delta_{\mu\nu}
+ F(2F-1)n_{\rm tot}\hat{f}_\mu \hat{f}_\nu,
\\
 \Psi_m^*(F_z^2 F_\mu + F_\mu F_z^2)_{mn}\Psi_n
&= Fn_{\rm tot}\left[
(2F-1)\hat{f}_z\delta_{\mu z}
+ (2F^2 - 3F +1)\hat{f}_z^2\hat{f}_\mu + F\hat{f}_\mu
\right].
\end{align}
Equation~(\ref{eq.b3}) is rewritten as
\begin{align}
 \mathrm{Re}\left[
\frac{\Psi_m^*(F_\mu)_{ml} H_{ln} \Psi_n}{Fn_{\rm tot}}
\right]
&= \frac{\Psi_m^*H_{mn}\Psi_n}{n_{\rm tot}} \hat{f}_\mu
+ \frac12 \left[ p\hat{z} + q(2F-1)\hat{f}_z\hat{z}
+ c_{\rm dd}\bm{b} \right]_\mu
\nonumber\\
&\quad
- \frac12 \left\{ \hat{\bm{f}}\cdot \left[ 
p\hat{z} + q(2F-1)\hat{f}_z\hat{z}
+ c_{\rm dd}\bm{b} \right] \right\}_\mu
\nonumber\\
&= \frac{\Psi_m^*H_{mn}\Psi_n}{n_{\rm tot}} \hat{f}_\mu
- \frac12 \left\{ \hat{\bm{f}}\times \left[
\hat{\bm{f}}\times \left(
p\hat{z} + q(2F-1)\hat{f}_z\hat{z} + c_{\rm dd}\bm{b}
\right) \right] \right\}_\mu,
\label{eq.Re_FH}
\end{align}
where
\begin{align}
\Psi_m^*H_{mn}\Psi_n
&= \Psi_m^*H_{mn}^{\rm s}\Psi_n + n_{\rm tot}\left\{
U_{\rm trap} + pF\hat{f}_z 
+ \frac{q}{2}F\left[ 1 + 2(F-1)\hat{f}_z^2 \right] 
+ c_{\rm dd}F\bm{b}\cdot\hat{\bm{f}}
\right\}.
\end{align}

Substituting Eqs.~(\ref{eq.Re_nabla2}), (\ref{eq.Im_nabla2}),
(\ref{eq.Im_FH}), and (\ref{eq.Re_FH}) into Eq.~(\ref{eq.dfdt.1}),
we obtain Eqs.~(\ref{eq.extLLG}) and (\ref{eq.B_eff}) as follows:
\begin{align}
 \frac{\partial \hat{\bm{f}}}{\partial t}
&= \frac{1}{1+\Gamma^2}\left[
\frac{1}{\hbar}\hat{\bm{f}} \times \bm{B}_{\rm eff}
- (\bm{v}_{\rm mass}\cdot\nabla)\hat{\bm{f}}
\right]
- \frac{\Gamma}{1+\Gamma^2}\hat{\bm{f}} \times\left[
\frac{1}{\hbar}\hat{\bm{f}} \times \bm{B}_{\rm eff}
- (\bm{v}_{\rm mass}\cdot\nabla)\hat{\bm{f}}
\right],
\\
\bm{B}_{\rm eff} &= \frac{\hbar^2}{2M}\nabla^2\hat{\bm{f}}
+\frac{\hbar^2}{2M}(\bm{a}\cdot\nabla)\hat{\bm{f}}
- c_{\rm dd}\bm{b} - p\hat{z}
- q(2F-1)\hat{f}_z\hat{z}.
\end{align}


\begin{thebibliography}{99} 
\bibitem{Leggett} A. J. Leggett, 
	Rev. Mod. Phys. {\bf 47}, 331 (1975).
\bibitem{Borovik} A. S. Borovik-Romanov, Y. M. Bunkov, V. V. Dmitriev,
	Y.M. Mukharskiy, D.A. Sergatskov,
	Phys. Rev. Lett. {\bf 62}, 1631 (1989).
\bibitem{Sadler2006}
	L. E. Sadler, J. M. Higbie, S. R. Leslie, M. Vengalattore, and D. M. Stamper-Kurn,
	Nature {\bf 443}, 312 (2006).
\bibitem{berkeley08} M. Vengalattore, S. R. Leslie, J. Guzman, and
	D. M. Stamper-Kurn, Phys. Rev. Lett. {\bf 100}, 170403 (2008).
\bibitem{Vengalattore2010} 
	M. Vengalattore, J. Guzman, S. R. Leslie, F. Serwane, and D. M. Stamper-Kurn, 
	Phys. Rev. A {\bf 81}, 053612 (2010).
\bibitem{lama} 
	A. Lamacraft, Phys. Rev. A {\bf 77}, 063622 (2008).
\bibitem{Kudo} K. Kudo and Y. Kawaguchi, 
	Phys Rev. A {\bf 82}, 053614 (2010).
\bibitem{Barnett2009}
	R. Barnett, D. Podolsky, and G. Refael,
	Phys. Rev. B {\bf 80}, 024420 (2009).
\bibitem{Cherng2011} R. W. Cherng and E. Demler, 
	Phys. Rev. A {\bf 83}, 053613 (2011); 
	R. W. Cherng and E. Demler, 
	Phys. Rev. A {\bf 83}, 053614 (2011).
\bibitem{LLG_ad} J. C. Slonczewski, 
	J. Magn. Magn. Mater. {\bf 159}, L1 (1996);
	L. Berger, 
	Phys. Rev. B {\bf 54}, 9353 (1996).
\bibitem{LLG_ad2} Ya. B. Bazaliy, B. A. Jones, and S.-C. Zhang, 
	Phys. Rev. B {\bf 57}, R3213 (1998); 
	Z. Li and S. Zhang, Phys. Rev. Lett. {\bf 92}, 207203 (2004).
\bibitem{Ho1996}
	T.-L. Ho and V. B. Shenoy,
	Phys. Rev. Lett. {\bf 77}, 2595 (1996).
\bibitem{Nakahara2000}
	M. Nakahara, T. Isoshima, K. Machida, S. Ogawa, and T. Ohmi,
	Physica B: Condensed Matter {\bf 284--288}, 17 (2000);
	T. Isoshima, M. Nakahara, T. Ohmi, and K. Machida,
	Phys. Rev. A {\bf 61}, 063610 (2000).
\bibitem{Mermin} N.D. Mermin and T.-L. Ho, Phys. Rev. Lett. {\bf 36},
	594 (1976).
\bibitem{LLGrev} M. Lakshmanan, Phil. Trans. R. Soc. A {\bf 369},
	1280-1300 (2011).
\bibitem{Tatara} G. Tatara, H. Kohno, and J. Shibata,
	Phys. Rep. {\bf 468}, 213 (2008).
\bibitem{LLG_nonad} S. Zhang and Z. Li, Phys. Rev. Lett. 
	{\bf 93}, 127204 (2004);
	A. Thiaville, Y. Nakatani, J. Miltat, and Y. Suzuki, 
	Europhys. Lett. {\bf 69}, 990 (2005).
\bibitem{Wong} 	C.H. Wong and Y. Tserkovnyak, 
	Phys. Rev. B {\bf 80}, 184411 (2009).
\bibitem{damp_micro} H. Kohno, G. Tatara, and J. Shibata, 
	J. Phys. Soc. Jpn. {\bf 75}, 113706 (2006);
	Y. Tserkovnyak, H.J. Skadsem, A. Brataas, G.E.W. Bauer, 
	Phys. Rev. B {\bf 74}, 144405 (2006).
\bibitem{LLG_theory} S. E. Barnes and S. Maekawa, 
	Phys. Rev. Lett. {\bf 95}, 107204 (2005);
	J. He, Z. Li, and S. Zhang, 
	Phys. Rev. B {\bf 73}, 184408 (2006).
\bibitem{shibata} J. Shibata, Y. Nakatani, G. Tatara, H. Kohno, and Y. Otani, 
	Phys. Rev. B {\bf 73}, 020403(R) (2006).
\bibitem{LLG_exp} A. Yamaguchi, et al., 
	Phys. Rev. Lett. {\bf 92}, 077205 (2004);
	{\it ibid.} {\bf 96}, 179904(E) (2006);
	M. Kl\"aui et al., 
	Phys. Rev. Lett. {\bf 95}, 026601 (2005).
\bibitem{Heyne} L. Heyne {\it et al.}, 
	Phys. Rev. Lett. {\bf 105}, 187203 (2010).
\bibitem{Ye} J. Ye, Y. B. Kim, A. J. Millis, B. I. Shraiman, 
	P. Majumdar, and Z. Te\v sanovi\'c, 
	Phys. Rev. Lett. {\bf 83}, 3737 (1999).
\bibitem{Onoda03} S. Onoda and N. Nagaosa, 
	Phys. Rev. Lett. {\bf 90}, 196602 (2003).
\bibitem{Onoda04} M. Onoda, G. Tatara, and N. Nagaosa, 
	J. Phys. Soc. Jpn. {\bf 73}, 2624 (2004).
\bibitem{Taguchi} K. Taguchi and G. Tatara,
	Phys. Rev. B {\bf 79}, 054423 (2009).
\bibitem{Kawaguchi2006}
	Y. Kawaguchi, H. Saito and M. Ueda,
	Phys. Rev. Lett. {\bf 96}, 080405 (2006);
	Phys. Rev. Lett. {\bf 97}, 130404 (2006).
\bibitem{Yi2006}
	S. Yi and H. Pu,
	Phys. Rev. Lett. {\bf 97}, 020401 (2006).
\bibitem{Landau} L. Landau and E. Lifshitz,
	Phys. Z. Sowjetunion {\bf 8}, 153 (1935).
\bibitem{magnetic_domains} A. Hubert and R. Sch\"{a}fer, 
	{\it Magnetic Domains}, (Springer-Ferlag, Berlin 1998).
\bibitem{Deutsch} J. M. Deutsch and T. Mai, 
	Phys Rev. E {\bf 72}, 016115 (2005).
\bibitem{Gerbier} F. Gerbier, A. Widera, S. Folling, O. Mandel, I. Bloch,
	Phys. Rev. A {\bf 73}, 041602(R) (2006). 
\bibitem{Tsubota} M. Tsubota, K. Kasamatsu, and M. Ueda, 
	Phys. Rev. A {\bf 65}, 023603 (2002); 
	K. Kasamatsu, M. Tsubota, and M. Ueda, 
	Phys. Rev. A {\bf 67}, 033610 (2003).
\bibitem{Choi} S. Choi, S.A. Morgan, and K. Burnett, 
	Phys. Rev. A {\bf 57}, 4057 (1998).
\bibitem{kawa09} Y. Kawaguchi, H. Saito, K. Kudo, and M. Ueda,
	Phys. Rev. A {\bf 82}, 043627 (2010). 
\bibitem{Ezawa} M. Ezawa,
	Phys. Rev. Lett. {\bf 105}, 197202 (2010).
\bibitem{Lin} Y.-J. Lin, R. L. Compton, K. Jim\'enez-Garc\'ia, 
	J. V. Porto, and  I. B. Spielman,
	Nature {\bf 462}, 628 (2009).

\end{thebibliography}
\end{document}